\documentclass[a4paper,fleqn,usenatbib]{mnras}

\usepackage[T1]{fontenc}
\usepackage{ae,aecompl}

\usepackage{graphicx}	
\usepackage{amsmath}	
\usepackage{amssymb}	
\usepackage{multicol}
\usepackage{bm}
\usepackage{pdflscape}
\usepackage{relsize}
\usepackage[normalem]{ulem}

\title[New Temperature and Metallicity Scale of Cool Giants]{\textbf{New Temperature and Metallicity Scale of Cool Giants from $K$-band Spectra}}
\author[Ghosh et al. 2019]{
Supriyo Ghosh,$^{1}$\thanks{E-mail: supriyo.ghosh@tifr.res.in (SG)},
D. K. Ojha$^{1}$, and 
J. P. Ninan$^{2,3}$ 
\\
$^{1}$Tata Institute of Fundamental Research, Homi Bhabha Road, Colaba, Mumbai-400 005, India \\
$^{2}$Department of Astronomy \& Astrophysics, 525 Davey Laboratory, The Pennsylvania State University, University Park, PA 16802, USA\\
$^{3}$ Center for Exoplanets and Habitable Worlds, 525 Davey Laboratory, The Pennsylvania State University,University Park, PA 16802, USA\\
}

\date{Accepted XXX. Received YYY; in original form ZZZ}

\pubyear{2019}

\begin{document}
\label{firstpage}
\pagerange{\pageref{firstpage}--\pageref{lastpage}}
\maketitle

\begin{abstract}
We present here quantitative diagnostic tools for cool giants that employ low-resolution near-infrared spectroscopy in the $K$-band for stellar population studies. In this study, a total of 260 cool giants (177 stars observed with X-shooter and 83 stars observed with NIFS) are used covering a wider metallicity range than in earlier works. We measure equivalent widths of some of the selected important $K$-band spectral features like Na I, Fe I and $^{12}$CO after degrading the spectral resolution (R$\sim$ 1200) to investigate the spectral behavior with fundamental parameters (e.g. effective temperature and metallicity). We derive empirical relations to measure effective temperature using the $^{12}$CO first-overtone band at 2.29 $\mu$m and 2.32 $\mu$m and show a detailed quantitative metallicity dependence of these correlations. We find that the empirical relations based on solar-neighborhood stars can incorporate large uncertainty in evaluating $T_{eff}$ for metal-poor or metal-rich stars. Furthermore, we explore all the spectral lines to establish the empirical relation with metallicity and find that the quadratic fit of the combination of Na I and $^{12}$CO at 2.29 $\mu$m lines yields a reliable empirical relation at [$Fe/H$] $\leq$ --0.4 dex, while a linear fit of any line offers a good metallicity scale for stars having [$Fe/H$] $\geq$ 0.0 dex.
\end{abstract}




\section{Introduction}

The estimation of fundamental parameters, e.g., effective temperature ($T_{eff}$), surface gravity (log $g$), metallicity ([$Fe/H$]), is very important to understand and classify stellar populations in different environments. Near-infrared (NIR) spectra, more precisely $K$-band (2.0--2.4 $\mu$m) spectral region, circumvent the problems of photometric as well as optical spectral measurements in the heavily reddened regions such as the Galactic bulge and Galactic plane. This is mainly because of a factor of 10 lower extinction in $K$-band than in $V$-band \citep{1989ApJ...345..245C} and the enhancement of contrast between brighter cluster giants and foreground field stars, often by as much as 3 to 5 mag \citep{2001AJ..122..1896}. Moreover, NIR $K$-band of cool giants ($T_{eff}$ $\leq$ 5000 K) offers very important diagnostic spectral features such as Na I doublet at 2.21 $\mu$m, the Ca I triplet at 2.26 $\mu$m, $^{12}$CO first-overtone bandhead at 2.29 $\mu$m (hereafter, CO229). The easiest and powerful approach to estimating parameters is implementing empirical correlations between observed line-strength indices and parameters. However, accurate, prior knowledge of the behavior of the spectral features with parameters in different stellar populations with broad parameter coverage is required for the precise characterization. 

Since the pioneering work of \citet{1970aj..75..785} and \citet{1986apjs..62..501}, many works have been done to investigate the sensitivity of the NIR spectral features of cool giants, especially in $K$-band, with their fundamental parameters (e.g. \citealt{1993aap..280..536, 1997apjs..111..445, 1997AJ..113..1411, 1998ApJ...508..397M, 2000aj..120..2089, 2000AJ....120..833, 2001AJ..122..1896, 2004apjs..151..387, 2006A&A...458..609D, 2011apj..741..108, 2013aap...549A.129, 2016aap..590A..6, 2019MNRAS.484.4619G}). These studies reveal that the $K$-band spectral features such as Na I, Ca I and CO are a good indicator of $T_{eff}$, log $g$ and [$Fe/H$] and can be used for luminosity classification as well. \citet{1997AJ..113..1411} first obtained a CO229 $-$ $T_{eff}$ relation with a residual scatter of 140 K. Subsequently, many empirical relations are established for more precise estimation using various features and continuum bandpasses (see, \citealt{2011apj..741..108, 2019MNRAS.484.4619G}) or adopting new indices to evaluate the band strength (see, \citealt{2003ApJ...597..323B, 2008aap..489..885}). \citet{2000AJ....120..833} and \citet{2001AJ..122..1896} obtained metallicity empirical relation of M-giants based on the equivalent widths (EWs) of three strong features in their $K$-band, namely Na I, Ca I and CO229 using moderate resolution (R $\sim$ 1300--4800) NIR spectra. Recently, \citet{2019MNRAS.484.4619G} found a remarkably tight relation between the EWs of CO229 and log $g$ using low-resolution NIR spectra (R $\sim$ 1200). In the past, $K$-band spectra are also efficiently measured detailed chemical signatures of red giant stars in the innermost regions of Milky Way Galaxy (see, \citealt{2015A&A...573A..14R, 2017AJ....154..239R}).  \citet{2016aap..590A..6} used low-resolution spectra to study behaviour of $T_{eff}$ and [$Fe/H$] with spectral indices for 20 Galactic bulge stars. \citet{2015ApJ...809..143D} and \citet{2017mnras..464..194} derived fundamental parameters of red giants stars in the nuclear star cluster and found that the majority of the stars is metal-rich. To summarize, we opine that the prominent $K$-band features in the NIR spectrum of cool stars and its potential to study the properties of stellar populations have been extensively acknowledged in the literature, and the empirical relations from these features are applied to characterize and classify the different stellar populations. Despite all the efforts, an additional study would be valuable to improve the quality and consistency of empirical relations suitable for stellar population studies. Moreover, the majority of prior work for empirical calibrations focusses on bright local solar neighborhood samples with a poor coverage of the atmospheric parameter space, especially in metallicity space. The poor metallicity coverage of previous spectral libraries was the limitation to explore the metallicity dependence of the spectral features in the $K$-band. In this context, the second data release of the X-shooter stellar library \citep{2014aap..565..117, 2020A&A...634A.133G} would be highly beneficial with a wider metallicity coverage (--2.5 < [Fe/H] < +1.0, \citealt{2019A&A...627A.138A}) than the previous libraries. 

The main motivation of this paper is, therefore, to provide an easy to use reliable empirical relations between fundamental parameters ($T_{eff}$ and [$Fe/H$]) and spectral line-strengths of cool giants. We make use of NIR $K$-band spectral features of cool giants covering a wide range of metallicities. The main advantages of empirical relations based on spectroscopically measured parameters are as follows. First, they yield accurate fundamental parameters of cool giants in different stellar populations by measuring only the line-strength of spectral features and second, they are independent of the reddening or distance to the object. Furthermore, we show the metallicity dependence of the spectral features, more importantly for a wider metallicity range than previous studies. In addition, this work evaluates how precisely the fundamental parameters such as $T_{eff}$ and [$Fe/H$] can be obtained from low-resolution $K$-band spectra. This would be highly valuable to understand the usefulness of low-resolution spectrographs for fundamental parameters estimation in stellar populations study. The paper is organized as follows. The sample giants are described in section~\ref{Data_collection} and section~\ref{Result_and_Discussion} deals with our new results and discussion. Finally, the summary of the work and conclusions are drawn in section~\ref{Summary_and_Conclusions}.   

\section{Sample selection} \label{Data_collection}
\begin{figure}
	\includegraphics[scale=0.55]{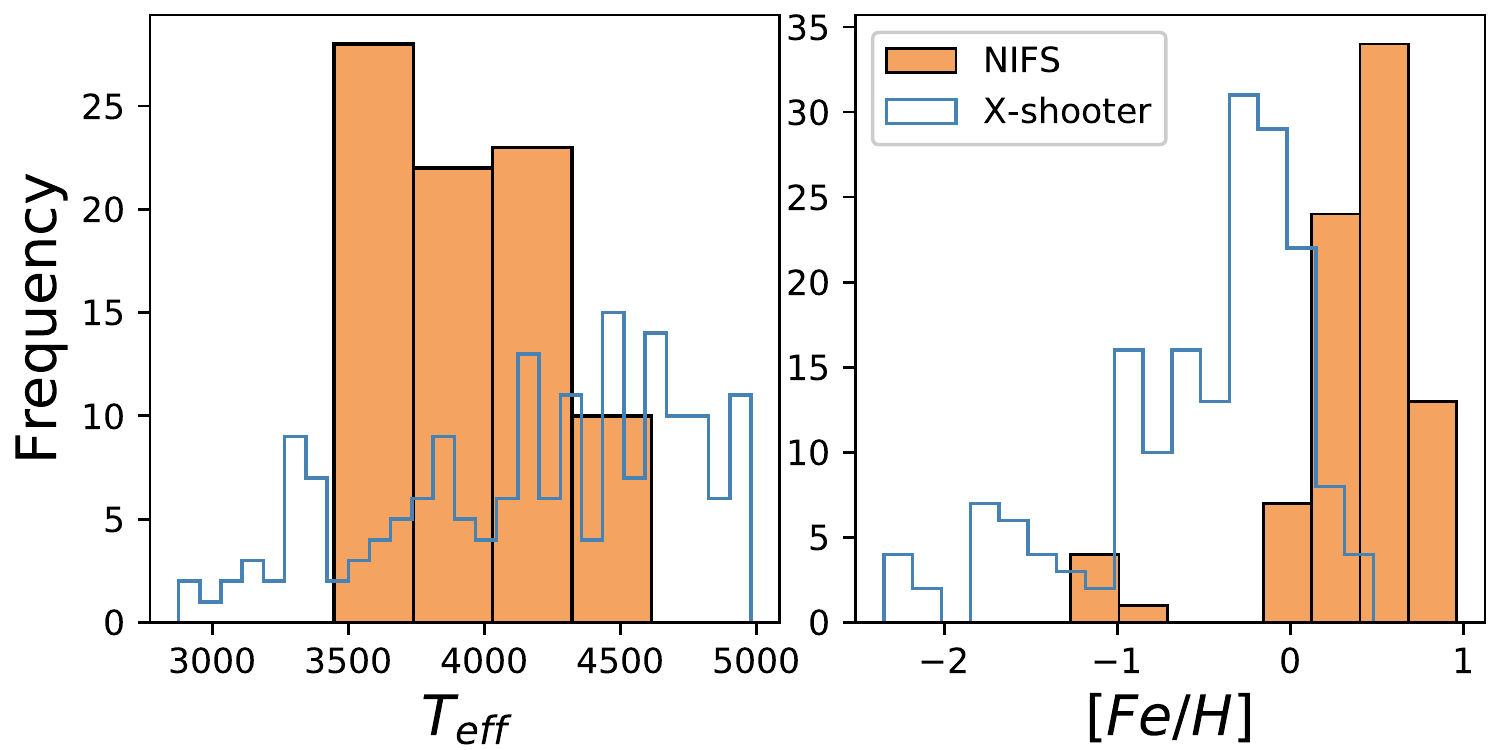}
    \caption{Histograms of the stellar parameters (Left: $T_{eff}$, and Right: [Fe/H]) for 260 (177 X-shooter, 83 NIFS) cool giants.}
    \label{Fig:sample_selection}
\end{figure}
In this work, we obtain near-infrared $K$-band spectra of 83 late-type giants, observed with the medium spectral resolution ($R$ $\sim$ 5400) Near-Infrared Facility Spectrograph (NIFS) on Gemini North within the central 1 pc of the Milky Way nuclear star cluster \citep{2015ApJ...808..106S, 2015ApJ...809..143D} and 381 giants having an effective temperature less than 5000 K from the X-shooter Spectral Library ($R$ $\sim$ 10000, the second data release, \citealt{2020A&A...634A.133G}) located in star clusters, in the field, in the Galactic bulge and in the Magellanic Clouds (we refer to \citealt{2020A&A...634A.133G} for details). The details about the instruments, observations and data reduction can be found in \citet{2015ApJ...809..143D} and \citet{2015ApJ...808..106S} for NIFS, and  \citet{2011A&A...536A.105V} and \citet{2020A&A...634A.133G} for X-shooter. We use SIMBAD to remove known supergiants, Mira variables and OH/IR stars of the X-shooter library from our study as they behave differently than normal giants \citep{2000aaps..146..217, 2018aj..155..216}.  The sample size reduces to 240 stars. Among them, 33 stars are observed more than once. Thus, our sample further reduces with 177 stars. We have obtained spectra of a total 260 (177 X-shooter, 83 NIFS) cool giants for this study.

 The $T_{eff}$ and [$Fe/H$] of these stars are taken from \citet{2015ApJ...809..143D} and \citet{2019A&A...627A.138A}. \citet{2015ApJ...809..143D} derived the parameters using spectral template fitting with the MARCS synthetic spectral grid \citep{2008A&A...486..951}. On the other hand, \citet{2019A&A...627A.138A} applied the full-spectrum fitting package University of Lyon Spectroscopic analysis Software (ULySS, \citealt{2009A&A...501.1269K}) with the Medium-resolution INT Library of Empirical Spectra (MILES) library \citep{2006MNRAS.371..703S, 2011A&A...532A..95F} as reference to fit the ultraviolet-blue and visible spectra for parameter estimation. Additional details about the fitting can be found in respective papers. For X-shooter stars with more than one observation, we use the straight mean of the various measurements. The precisions of the measurements are 400 K, 0.3 dex and 0.9 dex for NIFS stars and 26--132 K, 0.14--0.21 dex and 0.06--0.20 dex for X-shooter stars in $T_{eff}$, [$Fe/H$] and log $g$, respectively. The distribution of our sample in $T_{eff}$ and [$Fe/H$] space is shown in Figure~\ref{Fig:sample_selection} and the parameters of the sample stars are listed in Table~\ref{tab:all_measured_ews}. Our sample spans a wide range of $T_{eff}$ ($\sim$ 3000 to 5000 K) and [$Fe/H$] ($\sim$ --2.35 to +0.96) ensuring that we can explore possible empirical relations between spectral features and parameters for a wide range of metallicity and study possible metallicity dependence on those empirical relations.

\begin{table*}
	\centering
	\caption{Fundamental parameters and measured EWs of the sample.}
	\label{tab:all_measured_ews}
	\resizebox{0.75\textwidth}{!}{
	\begin{tabular}{crrrrrr} 
     \hline
Stars names &$T_{eff}$ &[$Fe/H$] &EW$_{NaI}$          &EW$_{FeI}$          &EW$_{CO229}$        &EW$_{CO232}$ \\
	\hline
 & & & \textbf{X-shooter} & & & \\
ISO-MCMS J004950.3--731116 & 3827 $\pm$ 52 & --0.52 $\pm$ 0.17 & 1.759 $\pm$ 0.503 & 0.981 $\pm$ 0.379 & 17.586 $\pm$ 1.784 & 14.162 $\pm$ 2.182 \\
ISO-MCMS J005059.4--731914 & 3806 $\pm$ 51 & --0.92 $\pm$ 0.17 & 1.756 $\pm$ 0.380 & 0.841 $\pm$ 0.286 & 19.269 $\pm$ 2.320 & 15.422 $\pm$ 2.641 \\
$[$M2002$]$ SMC 83593 & 3607 $\pm$ 59 & --0.98 $\pm$ 0.17 & 1.985 $\pm$ 0.482 & 0.67 $\pm$ 0.281 & 13.021 $\pm$ 1.391 & 10.765 $\pm$ 1.728 \\
ISO-MCMS J005314.8--730601 & 3762 $\pm$ 38 & --0.71 $\pm$ 0.09 & 2.034 $\pm$ 0.507 & 1.079 $\pm$ 0.331 & 19.865 $\pm$ 1.949 & 15.41 $\pm$ 2.303 \\
ISO-MCMS J005332.4--730501 & 4391 $\pm$ 32 & --0.58 $\pm$ 0.06 & 1.485 $\pm$ 0.358 & 0.753 $\pm$ 0.272 & 13.038 $\pm$ 1.017 & 10.968 $\pm$ 1.828 \\
SHV 0549503--704331 & 3089 $\pm$ 51 & --0.38 $\pm$ 0.17 & --0.426 $\pm$ 0.143 & --0.319 $\pm$ 0.109 & 3.025 $\pm$ 1.013 & --0.706 $\pm$ 0.643 \\
HV 2360 & 3352 $\pm$ 34 & --0.64 $\pm$ 0.09 & 3.221 $\pm$ 0.632 & 1.552 $\pm$ 0.529 & 18.538 $\pm$ 2.224 & 14.801 $\pm$  2.433 \\
HV 2446 & 2876 $\pm$ 35 & --0.21 $\pm$ 0.13 & 3.048 $\pm$ 0.558 & 2.342 $\pm$ 0.856 & 18.505 $\pm$ 2.050 & 13.655 $\pm$ 2.556 \\

...     &...       &...    &...             &...                 &...                 &...                 \\
		\hline
\end{tabular}	}

      \small
      Table~\ref{tab:all_measured_ews} is available in its entirety in the electronic version of the journal as supplementary material.
\end{table*}

\begin{table*}
\centering
\caption{Spectral bands for EWs estimation.}
\label{tab:bandpass}

\resizebox{0.75\textwidth}{!}{
\begin{tabular}{lcccr} 
   \hline
   \hline
Index & Feature & Feature  & Continuum  &  Ref. \\	 
      &         & Bandpass ($\mu$m) & Bandpass ($\mu$m) &  \\
\hline

NaI & Na~I (2.21 $\mu$m)  & 2.2040--2.2107 & 2.1910--2.1966, 2.2125--2.2170 & 1\\
FeI & Fe~I (2.23 $\mu$m) & 2.2250--2.2299 & 2.2133--2.2176, 2.2437--2.2479 & 2 \\
CO229 & $^{12}$CO(2-0) (2.29 $\mu$m) & 2.2910--2.3020 & 2.2420--2.2580, 2.2840--2.2910 & 3 \\
CO232 & $^{12}$CO(3-1) (2.32 $\mu$m) & 2.3218--2.3272 & 2.2325--2.2345, 2.2873--2.2900 & 4 \\
\hline
\end{tabular}
} 
\\
Ref : (1) \citet{2001AJ..122..1896}; (2) \citet{2013aap...549A.129}; (3) \citet{2019MNRAS.484.4619G} (4) \it{This work} 
\end{table*}

\begin{figure}
	\includegraphics[scale=0.60]{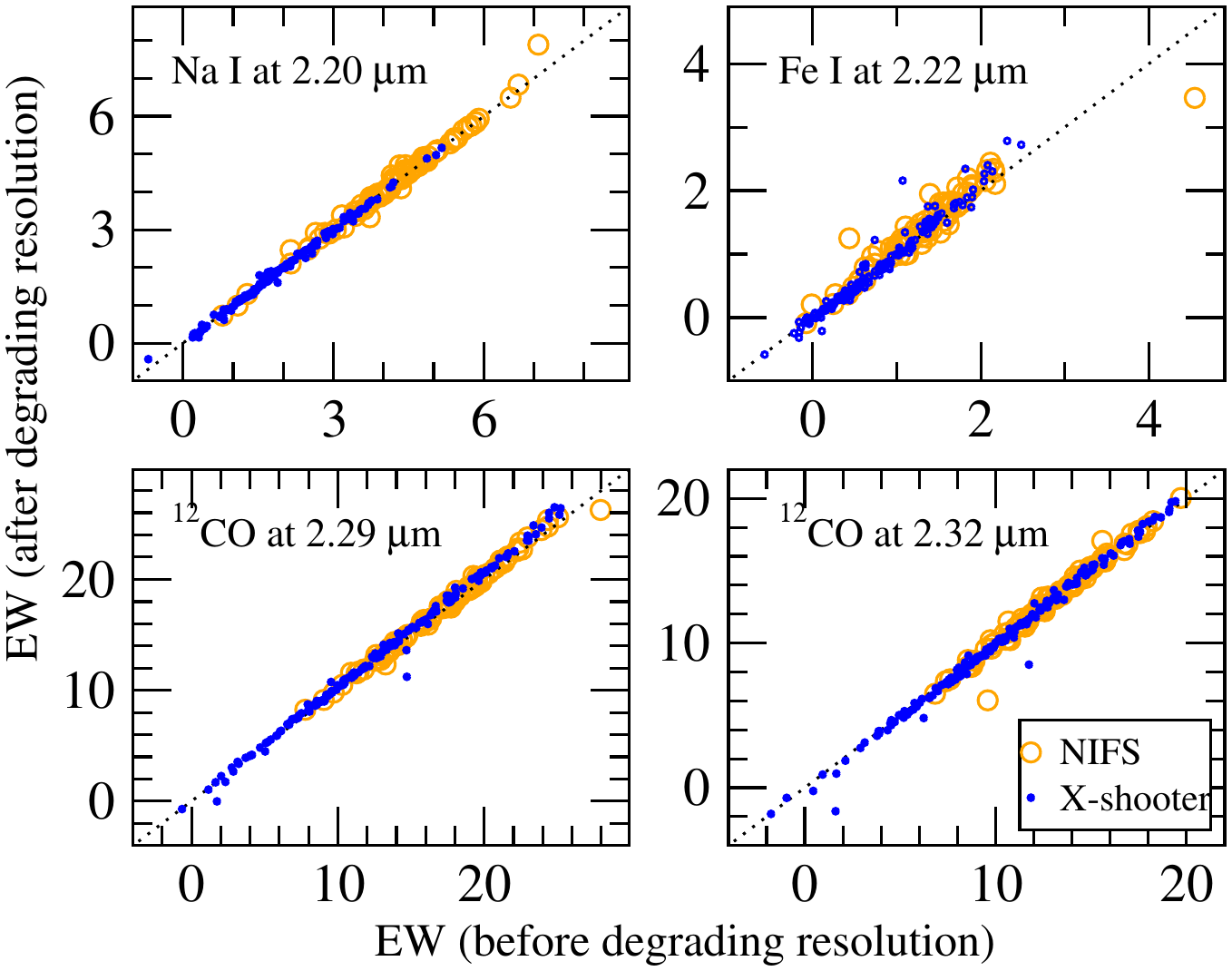}
    \caption{Study of the resolution effect on equivalent widths computation using the NIFS (from R$\sim$ 5400 to R$\sim$ 1200) and X-shooter (from R$\sim$ 10000 to R$\sim$ 1200) spectra. Orange and blue symbols display NIFS and X-shooter spectra, respectively. The dotted line displays the one-to-one correspondence of equivalent widths.}
    \label{Fig:resolution_effect}
\end{figure}

In this work, we estimate the strength of spectral features by measuring equivalent widths (EWs). The EWs of Na I at 2.20 $\mu$m (hereafter, NaI), Fe I at 2.22 $\mu$m (hereafter, FeI), $^{12}$CO (2--0) at 2.29 (hereafter, CO229) and $^{12}$CO (3--0) at 2.32 (hereafter, CO232) and its' uncertainties are estimated following the method as in \citet{2014AJ....147...20N}. The adopted continuum and feature bandpasses are listed in Table~\ref{tab:bandpass}. To estimate EWs, feature band and continuum bands of NaI and CO229 are adopted from \citet{2001AJ..122..1896} and \citet{2019MNRAS.484.4619G}, respectively. We compute the FeI line strength adopting the bandpass from \citet{2013aap...549A.129}. The CO232 feature has the central bandpass overlapping with the \citet{2019MNRAS.484.4619G} definition, whereas the continuum bandpasses (2.2325--2.2345 $\mu$m and 2.2873--2.2900 $\mu$m) are different. We select the different continuum as some narrow spikes sometimes arise near 2.25 $\mu$m in X-Shooter spectra \citep{2020A&A...634A.133G}, affecting the continuum of the \citet{2019MNRAS.484.4619G}. These narrow spikes also affect the feature band of the Ca I at 2.26 $\mu$m. Thus, we do not consider Ca I line in this study. Our main goal in this work is to study the spectral behavior at low-resolution, which helps to understand how precise stellar parameters can be evaluated from the spectra of low-resolution spectrographs like TIRSPEC (R $\sim$ 1200, \citealt{2014JAI.....350006}). Therefore, all the spectra are degraded to TIRSPEC spectral resolution before computing EWs to eliminate possible resolution effect, and the spectral features are corrected for the zero velocity by shifting. The measured EWs are listed in Table~\ref{tab:all_measured_ews}. Although we degrade the resolution of all the spectra,  the resolution effect on EWs computation is investigated using the NIFS and X-shooter spectra. A comparison of EWs before and after degrading resolution is presented in Fig.~\ref{Fig:resolution_effect}. For NIFS, the mean and standard deviation of EWs before (after) degrading resolution are 4.17\AA~and 1.14\AA~(4.21\AA~and 1.18\AA) for NaI, 1.32\AA~and 0.61\AA~(1.41\AA~and 0.57\AA) for FeI, 17.15\AA~and 4.05\AA~(17.37\AA~and 4.11\AA) for CO229, 12.90\AA~and 2.79\AA~(12.92\AA~and 2.97\AA) for CO232, respectively. For X-shooter, the same parameters are 1.94\AA~and 1.06\AA~(1.93\AA~and 1.06\AA) for NaI, 0.76\AA~and 0.58\AA~(0.77\AA~and 0.65\AA) for FeI, 12.93\AA~and 5.90\AA~(13.31\AA~and 6.09\AA) for CO229, 10.36\AA~and 4.36\AA~(10.28\AA~and 4.53\AA) for CO232, respectively. This test shows that overall, degrading the resolution shows no significant impact on EWs computation.

\section{Result and Discussion} \label{Result_and_Discussion}
 
\begin{figure*}
	\includegraphics[scale=0.75]{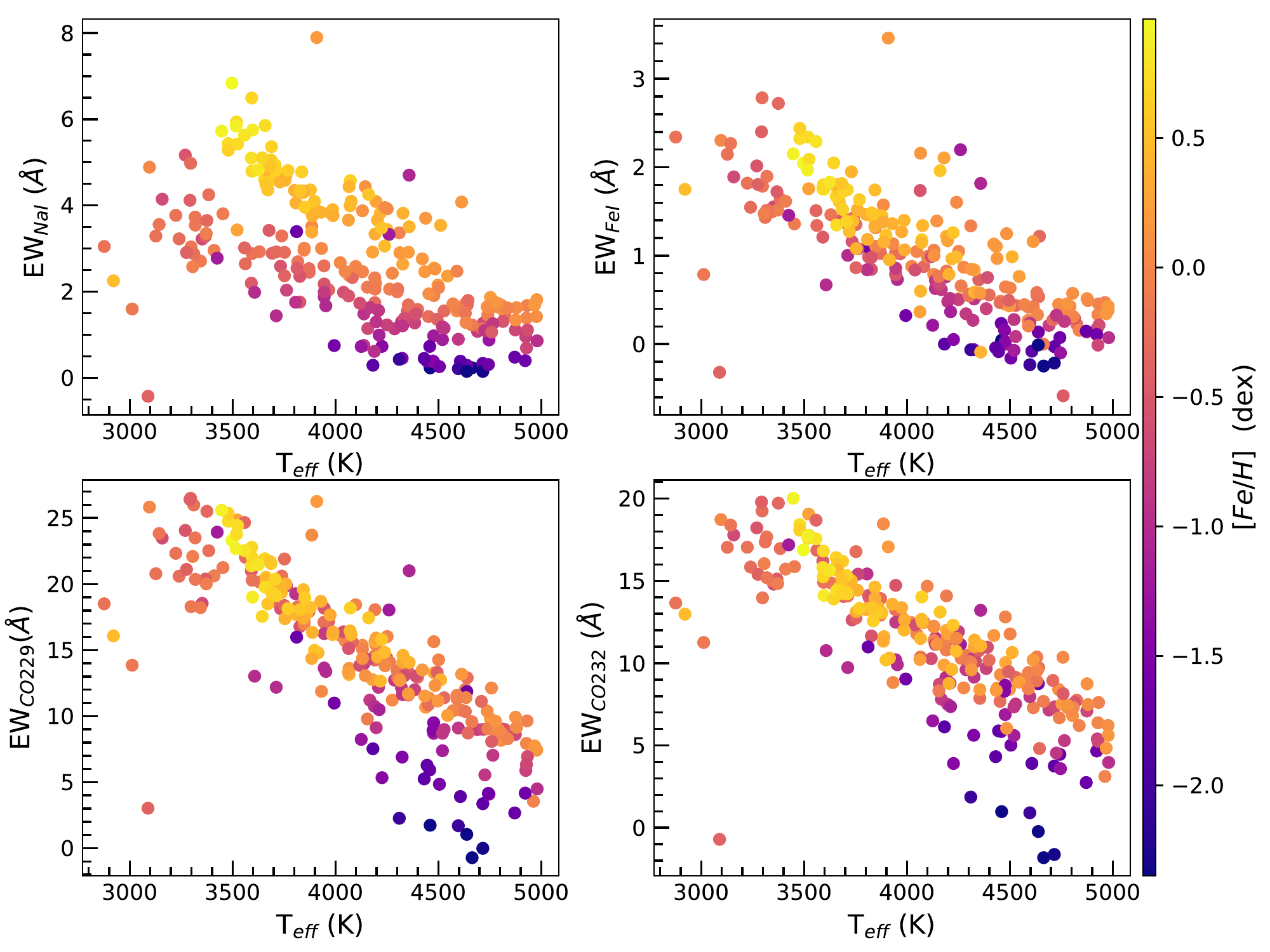}
    \caption{Variations of the equivalent widths of the lines corresponding to the $K$-band atomic and molecular absorption features as a function of effective temperature discussed in the text. The colour bar represents the metallicity of each star.}
     \label{Fig:index-Teff_behaviour}
\end{figure*}
\begin{figure}
	\includegraphics[scale=0.635]{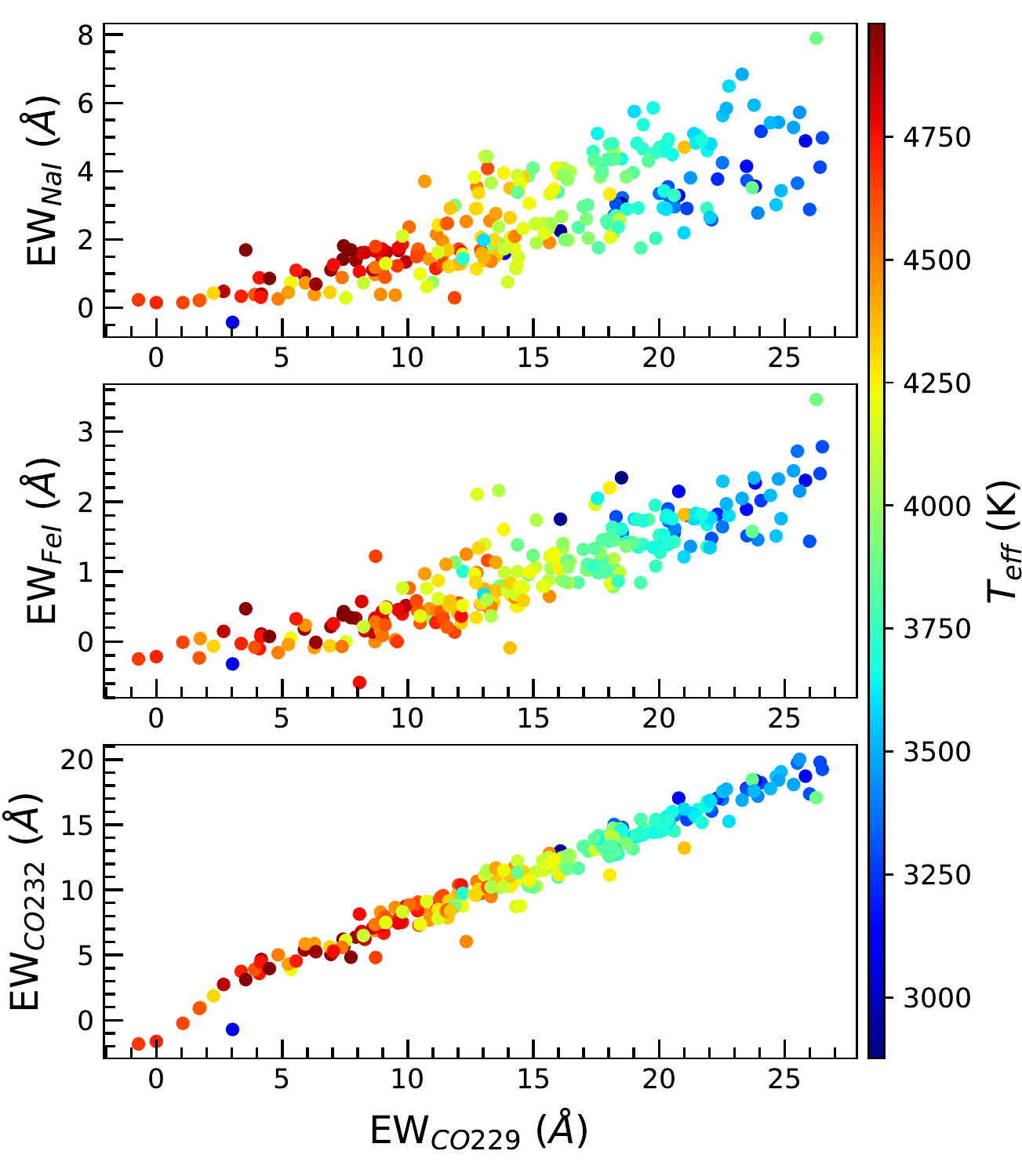}
    \caption{Diagnostic diagrams for investigating the origin of the dispersion in especially Figure~\ref{Fig:index-Teff_behaviour} by index-index plot. The colour bar represents the effective temperature of each star.}
    \label{Fig:index-index_relation}
\end{figure}
\begin{figure*}
	\includegraphics[scale=0.75]{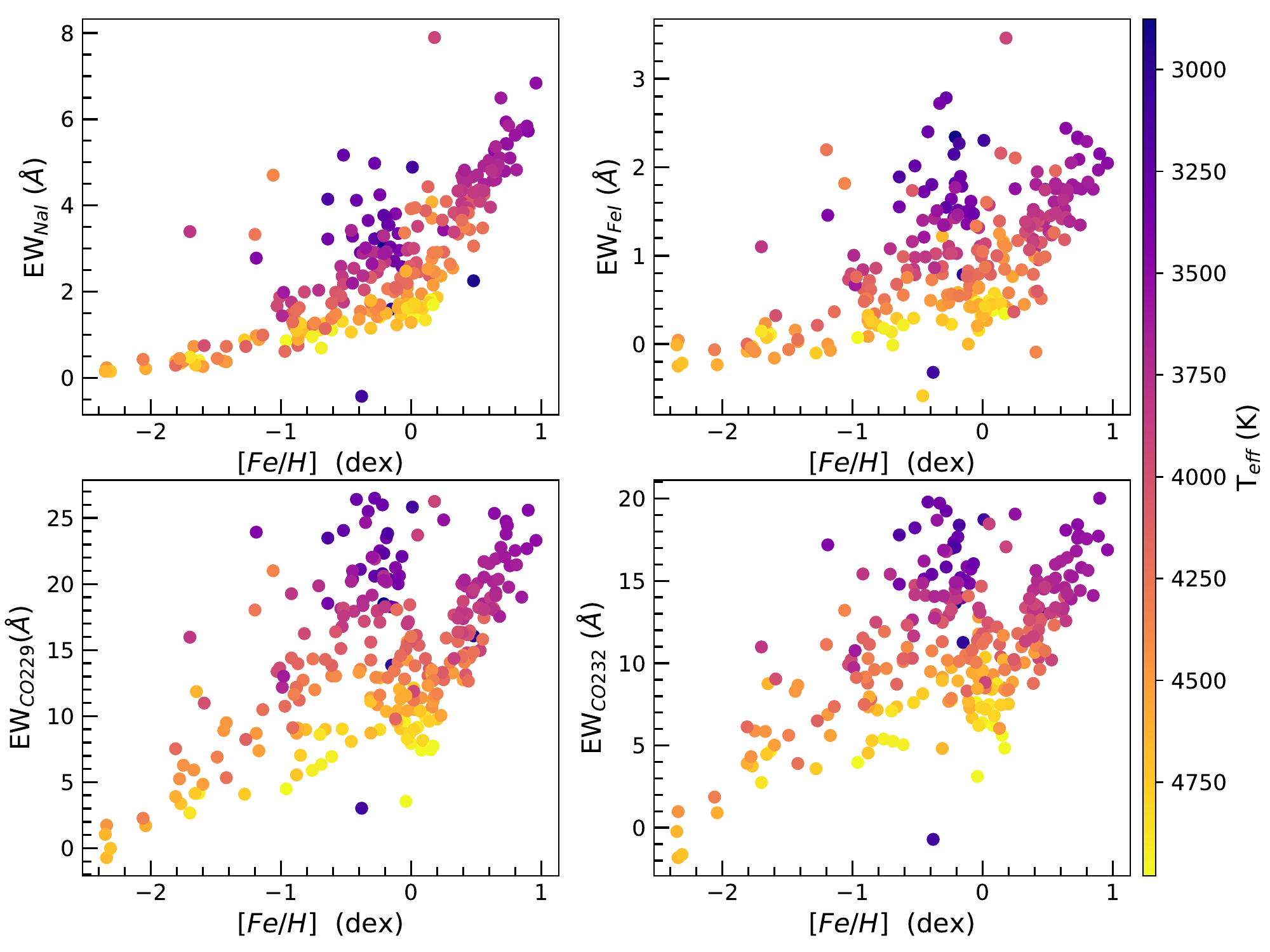}
    \caption{Variations of the equivalent widths of the lines corresponding to the $K$-band atomic and molecular absorption features as a function of metallicity discussed in the text.  The colour bar represents the effective temperature of each star.}
    \label{Fig:index-meta_behaviour}
\end{figure*}

\subsection{Behaviour of selected features with stellar parameters}
To study the behaviour of the spectral features with the parameters ($T_{eff}$ and [$Fe/H$]), we select most prominent atomic (NaI and FeI) and molecular (CO229 and CO232) signatures of $K$-band spectra. The behaviour of those lines with $T_{eff}$ and [$Fe/H$] is shown in Fig.~\ref{Fig:index-Teff_behaviour} and Fig.~\ref{Fig:index-meta_behaviour}. 

 As can be seen from Fig.~\ref{Fig:index-Teff_behaviour}, the strengths of all the absorption features of our interest (NaI, FeI, CO229, and CO232) strongly depend on $T_{eff}$ and show an increasing trend with decreasing $T_{eff}$ as found by previous studies (see, \citealt{2019MNRAS.484.4619G} and references therein). However, they display a large scatter. This is presumably due to the metallicity effect, mainly because of the large range in [Fe/H] covered by our sample stars. NaI shows more dispersion than other lines representing its' higher sensitivity on metallicity. For NaI, two distinct sequences in terms of [$Fe/H$] can be seen in Fig.~\ref{Fig:index-Teff_behaviour}. The upper sequence mainly contains metal-rich stars and lower sequence is for metal-poor stars. It is to be noted that our sample of metal-rich giants are located in the Galactic center (GC) and these GC stars show stronger NaI absorption than in the solar neighborhood giants. \citep{1996aj..112..1988, 2011apj..741..108}. The stronger line strength may be due to the increased rotational mixing in dense stellar clusters \citep{2011apj..741..108}. Furthermore, at low temperatures (for cooler stars than K3 giants), NaI lines are blends of a couple of atomic lines (e.g. Sc, Si, Fe and CN) as found by \citet{1996apjs..107..312} using high-resolution spectra. On the other hand, CO229 and CO232 lines show larger scatter for those stars having (i) $T_{eff}$ $>$ 4000 K and [$Fe/H$] $<$ --1.5 dex; and (ii) $T_{eff}$ $<$ 3400 K and [$Fe/H$] $\sim$ 0.0 dex. At higher temperature ($T_{eff}$ $>$ 4000 K), the dispersion may be caused by the metallicity, however, most of the giants become variable at a lower temperature ($T_{eff}$ $<$ 3400 K) and the variability of stars plays a significant role in dispersion. 
 
 We further investigate the origin of the dispersion in Fig.~\ref{Fig:index-Teff_behaviour} by plotting index-index correlation as depicted in Fig.~\ref{Fig:index-index_relation}.  It is expected a tight index-index correlation, especially in the case of CO-indices, which are, most likely, strongly correlated.  If an index-index relation is not so tight, this might be caused by varying abundance ratios, remains of telluric lines, etc. In our case, a very tight correlation is evident for CO232--CO229, but NaI--CO229 correlation shows a large scatter. This confirms that the large dispersion in Fig.~\ref{Fig:index-Teff_behaviour} is because of the large coverage in metallicity space by our sample stars.
 
 A variation of the EWs with metallicities is also evident in Fig.~\ref{Fig:index-meta_behaviour}, with an increase from low to high [Fe/H]. An increased dispersion of EWs or even a plateau can be found at about [Fe/H] $\leq$ 0.4 dex up to solar metallicity indicating the saturation of spectral lines. In addition to the decrease in effective temperature, the increase of metallicity is responsible for line saturation. Two distinct sequences can be seen in EW--[Fe/H] plane --- one is from sub-solar to solar and the other one is from solar to super-solar metallicity. A few sub-solar stars ([$Fe/H$] -- 0.0 to 1.0) with $T_{eff}$ $\geq$4500 K follow the solar to super-solar metallicity sequence. This behavior is very puzzling and could be because of the different abundance ratios in relatively warmer stars.

\subsection{Empirical relations}

\subsubsection{Effective temperature indicator} \label{Effective_temperature_correlation}
Two bandheads CO229 and CO232 are undertaken for new empirical relations with $T_{eff}$ and to inspect metallicity dependence of those relations. To establish empirical relations, we follow equations
\begin{equation} \label{equation:linear_individuallines}
z = m0 + a\times x
\end{equation}
for a linear fit of a individual line, and
\begin{equation} \label{equation:linear_combinedlines}
z = m0 + a\times x + b\times y
\end{equation}
for a linear fit of a combination of two lines, where z= fundamental parameter (e.g., $T_{eff}$), x and y are EWs of spectral features, m0, a and b are the coefficients of the fit.
As CO lines vary almost linearly with $T_{eff}$ (see, Fig.~\ref{Fig:index-Teff_behaviour}), a linear fit (using Eq.\ref{equation:linear_individuallines}) is explored for each bandhead separately after eliminating 2$\sigma$ outliers. The correlation coefficient ($R$), the coefficient of determination ($R_{sqr}$), and the standard error of estimate (SEE) are listed in Table~\ref{tab:Goodness of fit for Teff}. Four different cases are exercised to establish new empirical relations and to investigate a possible metallicity dependence in the $T_{eff}$--CO empirical relations as follows. 

\begin{figure*}
	\includegraphics[scale=0.45]{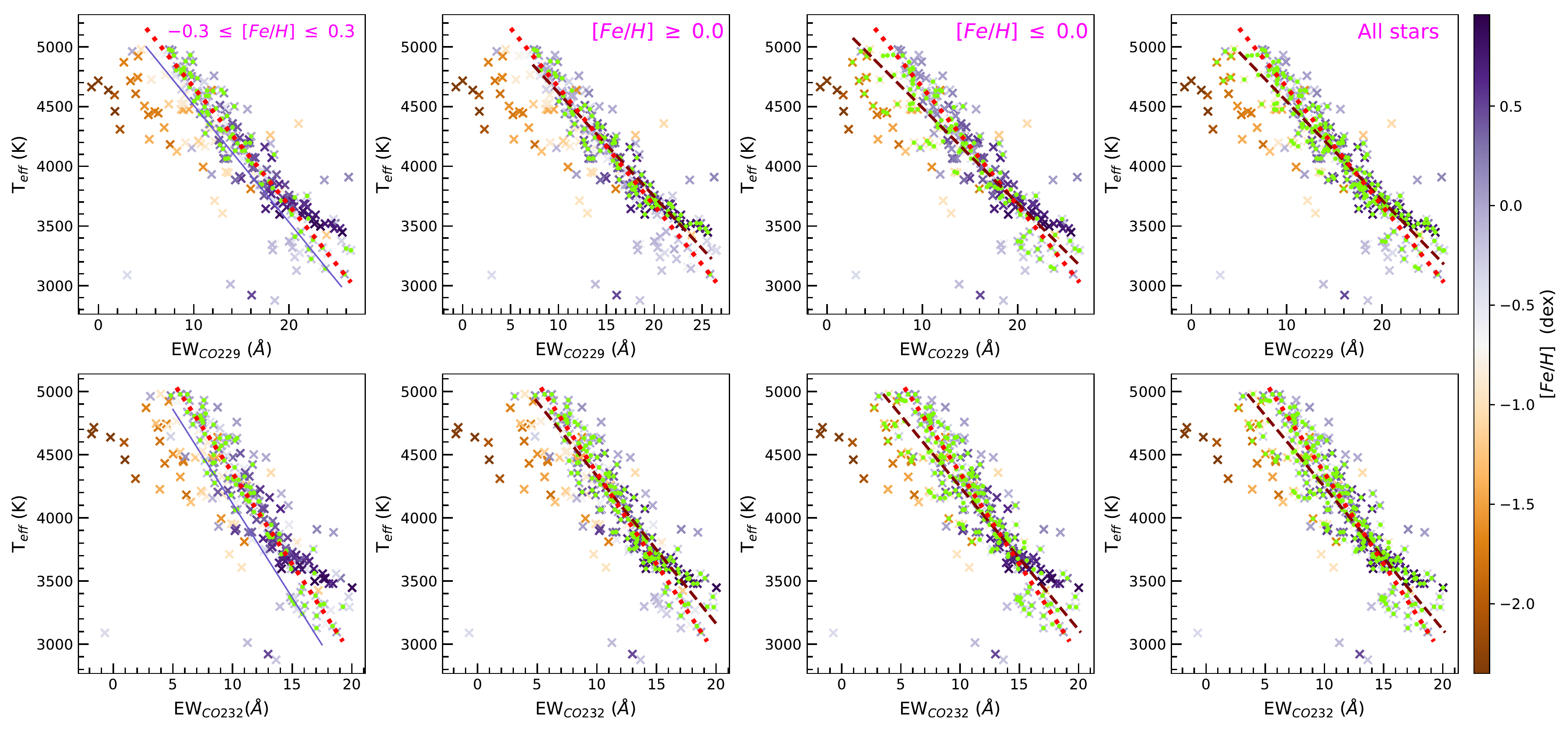}
    \caption{Empirical correlation between $T_{eff}$ and equivalent widths of $^{12}$CO (upper panel: CO at 2.29 $\mu$m, lower panel: CO at 2.32 $\mu$m). The colour bar represents the metallicity of each star. The colored `X' symbols refer to the whole sample and green dots represent the stars used to establish empirical relations after removing the 2$\sigma$ outliers in respective metallicity range. The red dot line indicates the best-fit relation for the stars belonging to the metallicity range --0.3 to +0.3 dex (Case 1) and the maroon dash line corresponds to the linear fit relation for others metallicity range (Case 2, Case 3 and Case 4). For comparison, we overplot Case 1 empirical relation with other cases. The blue line represents the empirical relation of \citet{2019MNRAS.484.4619G}, which was established using 107 solar-neighborhood giants.}
    \label{Fig:index-Teff_correlation}
\end{figure*}

\begin{figure*}
	\includegraphics[scale=0.55]{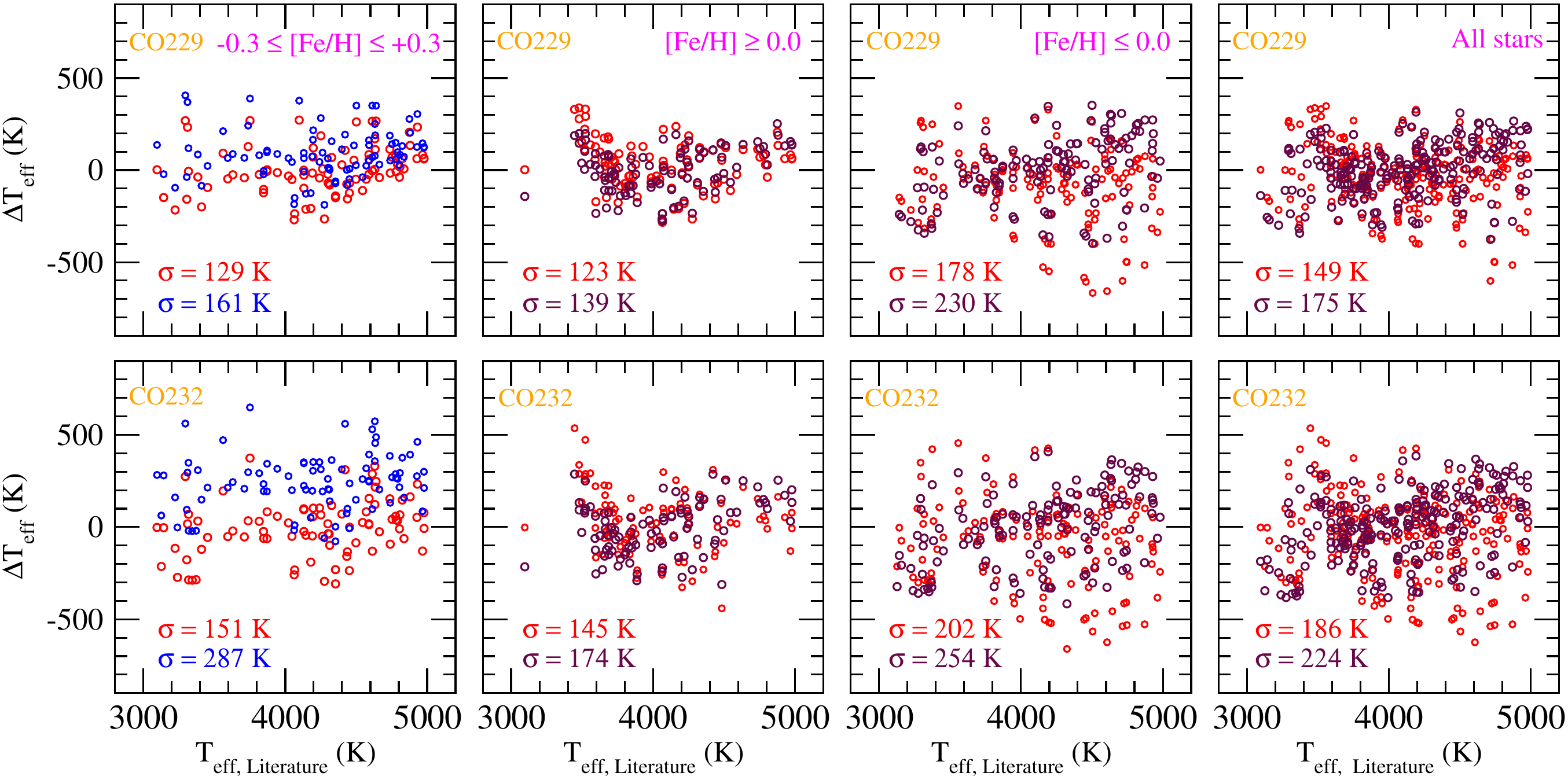}
    \caption{Residuals of the derived effective temperature (literature minus derived) from different established relations against the literature effective temperature are plotted for comparison. Here, we display only those stars that are considered for the empirical relation after 2$\sigma$ clipping. The red circles in the corresponding metallicity range represent the estimation applying the empirical relation of metallicity range from --0.3 to +0.3 dex. The blue circles represent the estimation using the empirical relation of \citet{2019MNRAS.484.4619G} and the maroon circles refer to the estimation using the empirical relations in the corresponding metallicity range.}
    \label{Fig:index-Teff_correlation_residual}
\end{figure*}

First (Case 1), we consider all the giants in our sample belonging to the metallicity range between --0.3 and +0.3 dex (considered here as solar-neighborhood stars) to minimize any potential metallicity effect on the empirical relation. The SEE of the fit is 128 K (153 K) for CO229 (CO232), which is comparable with the SEE of \citet{2019MNRAS.484.4619G}. The $T_{eff}$ vs. EWs plot for the sample stars is depicted in Fig.~\ref{Fig:index-Teff_correlation}. The colored `X' symbols refer to the whole sample and green dots represent the stars used to establish empirical relations after removing the 2$\sigma$ outliers. The red dot line indicates the best-fit relation for the stars belonging to the metallicity range --0.3 to +0.3 dex. The blue line represents the empirical relation from \citet{2019MNRAS.484.4619G}, which was established using 107 solar-neighborhood giants. As can be seen from Fig.~\ref{Fig:index-Teff_correlation}, the slopes of the two empirical relations are significantly different for both CO229 and CO232. The offset between empirical relations could be due to the different methods used to estimate the atmospheric parameters of the sample stars. While the atmospheric parameters of \citet{2019MNRAS.484.4619G} sample stars are derived by \citet{2017MNRAS.471..770M} by comparing multiwavelength archival photometry to BT-Settle model atmospheres, the full-spectrum fitting (see, \citealt{2015ApJ...809..143D, 2019A&A...627A.138A}) is applied to derive the parameters of the stars used in this work. Comparing the $T_{eff}$ obtained from both empirical relations, we find that the $T_{eff}$s are on average $\sim$120 K ($\sim$ 200 K) warmer for CO229 (CO232), respectively, than \citet{2019MNRAS.484.4619G}. We then inverted the process and estimated $T_{eff}$ for each of the stars using the empirical relations established in this work and \citet{2019MNRAS.484.4619G}. We show a comparison of the obtained values to the literature values in Fig.~\ref{Fig:index-Teff_correlation_residual}. The mean and standard deviation of the fit residuals are $\Delta T_{eff, Avg}$ = 4 K (--3 K), $\sigma_{T_{eff}}$ = 129 K (151 K) and  $T_{eff}$ = 93 K (247 K), $\sigma_{T_{eff}}$ = 161 K (287 K) for CO229 (CO232), respectively. It is to be noted that we only consider the stars that are fitted for the empirical relations to evaluate the mean and the standard deviation. It is expected that the inclusion of outlier stars would give a larger value of those parameters. 
 
Second (Case 2), we consider all the giants in our sample having metallicity $\geq$ 0.0 dex (metal-rich stars) and the best fit is displayed in Fig.\ref{Fig:index-Teff_correlation}. The maroon dash line refers to the linear fit. The number of the stars used for the fit after 2$\sigma$ clipping and the coefficients of fit are listed in Table~\ref{tab:Goodness of fit for Teff} along with SEE. The SEE of the fit is 124 K (146 K) for CO229 (CO232), respectively. For a comparison, the empirical relation of Case 1 is also overplotted in Fig.\ref{Fig:index-Teff_correlation} (red dotted line). The empirical relations of Case 1 and Case 2 are in good agreement only in a small regime of $T_{eff}$ (4500--4000 K). The effective temperature tends to be underestimated by up to $\sim$ 150 K (250 K) at $T_{eff}$ $\leq$ 4000 K for CO229 (CO232), respectively, but rather overestimated by up to $\sim$ 75 K ($\sim$ 160 K) at $\geq$ 4500 K if we estimate $T_{eff}$ using empirical relations established in Case 1. The different slope of the empirical relations indicates the metallicity dependence on $T_{eff}$--CO relation for metal-rich stars. We further observe that the majority of the stars belonging to the [$Fe/H$] $\geq$ 0.3 dex has $T_{eff}$ < 4000 K and those stars shift to warmer temperatures than their solar metallicity counterparts. This shift is caused by the increase of mean molecular weight at metallicities higher than about solar (see, \citealt{1998A&A...335..573M} for a review). Now, the two linear empirical solutions (Case 1 and Case 2) are applied to each star (excluding outliers) in the sample, and the resulting $T_{eff}$ values are compared with literature $T_{eff}$ in Fig.~\ref{Fig:index-Teff_correlation_residual}.  The mean and standard deviation of the fit residuals are $\Delta T_{eff, Avg}$ = 31 K (38 K), $\sigma_{T_{eff}}$ = 139 K (174 K) and  $T_{eff}$ = 1 K (-4 K), $\sigma_{T_{eff}}$ = 123 K (145 K) for Case 1 and Case 2, respectively.  

Third (Case 3), all the giants in the sample having metallicity $\geq$ 0.0 dex ( 0.0--2.35) are considered (metal-poor stars). The best fit is displayed in Fig.~\ref{Fig:index-Teff_correlation} by the maroon dash line and all the fit parameters are listed in Table~\ref{tab:Goodness of fit for Teff}. The SEE of the fit is 180 K (204 K) for CO229 (CO232), respectively. We believe that the larger SEE than in Case 1 is because of the dispersion caused by the large metallicity coverage (--1.81 to 0.0 after the fit removing 2$\sigma$ outliers) of the sample. For a comparison, the empirical relation of Case 1 is also overplotted in Fig.~\ref{Fig:index-Teff_correlation}. We find that the effective temperature tends to be underestimated by up to $\sim$ 250 K (260 K) at $T_{eff}$ $\geq$ 3800 K for CO229 (CO232), respectively, but rather overestimated by up to $\sim$ 140 K ($\sim$ 200 K) at $\leq$ 3800 K in Case 3 in comparison to $T_{eff}$ estimated using the empirical relations of Case 1. The large deviation of $T_{eff}$ represents the metallicity dependence CO--$T_{eff}$ empirical relations. Now, we estimate $T_{eff}$ to each star (excluding outliers) in the sample using the two linear empirical solutions (Case 3 and Case 1), and compare with literature $T_{eff}$ as shown in Fig.~\ref{Fig:index-Teff_correlation_residual}.  The mean and standard deviation of the fit residuals are $\Delta T_{eff, Avg}$ = --95 K (-- 87 K), $\sigma_{T_{eff}}$ = 230 K (254 K) and  $T_{eff}$ = 1 K (--4 K), $\sigma_{T_{eff}}$ = 178 K (202 K) for Case 1 and Case 3, respectively.

Fourth (Case 4), we use all the giants in the sample for the empirical relation. As can be seen from Fig.~\ref{Fig:index-Teff_correlation}, the best fit is displayed by the maroon dash line, and the empirical relation of Case 1 (red dot line) is overplotted for comparison. All the fit parameters are listed in Table~\ref{tab:Goodness of fit for Teff}. The SEE of the fit is 149 K (186 K) for CO229 (CO232), respectively. Similar to the Case 2 and the Case 3, the effective temperature tends to be underestimated by up to $\sim$ 180 K (240 K) at $T_{eff}$ $\geq$ 4000 K for CO229 (CO232), respectively, but rather overestimated by up to $\sim$ 150 K ($\sim$ 220 K) at $\leq$ 4000 K in Case 4 than the obtained $T_{eff}$ using the empirical relations of Case 1. In fact, the Case 4 relation can be considered as a combined effect of the Case 2 and Case 3, where the cooler end and the warmer end of the empirical relation follow the metal-rich and metal-poor stars, respectively. We then inverted the process and estimated $T_{eff}$ for each of the stars using the empirical relations of Case 4 and Case 1. We show a comparison of the obtained value to the literature value in Fig.~\ref{Fig:index-Teff_correlation_residual} with $\sigma_{T_{eff}}$ = 149 K (175 K) in Case 4 and $\sigma_{T_{eff}}$ = 186 K (224 K) in Case 1 for CO229 (CO232) line, respectively. The SEE of two relations differs because of the metallicity dependence on the empirical relation, where more metal-rich and metal-poor stars significantly deviate from the relation that uses a narrow metallicity range (i.e. Case 1).

Different case study unveils the variation of SEE and fit parameters of empirical relations that confirms the significant influence of metallicity on $T_{eff}$--CO correlation. However, \citet{2016aap..590A..6} did not find any metallicity dependence on $T_{eff}$--CO relation within the metallicity range --1.2 < [Fe/H] < 0.5 with a sample containing only 20 Galactic bulge stars (3 stars having [$Fe/H$] > 0 dex). We here showed with a larger sample that the empirical relations based on solar-neighborhood stars can incorporate large uncertainty in evaluating $T_{eff}$ for metal-poor or metal-rich stars. To decide which one among the 4 relations established in this work should be implemented to the unknown sample for $T_{eff}$ estimation certainly depends on whether we have previous knowledge of metallicity or not. If we know the metallicity of the stars, we can choose the empirical relation depending on the metallicity. Otherwise, the relation of Case 4 can be applied for $T_{eff}$ estimation in general with a typical accuracy of $\sim$ 150 K ($\sim$ 190 K) for CO229 (CO232) in the metallicity range from --1.81 dex to +0.96 dex and could be used reliably for metal-poor or metal-rich stars.  In addition, we investigate the empirical relation of the metallic lines like NaI and FeI with $T_{eff}$. However, the intrinsic scatter is much higher than CO--$T_{eff}$ and so we do not discuss those relations further. It also indicates the greater sensitivity of CO lines with $T_{eff}$ than metallic lines studied here.
 
 \begin{table*}
\centering
\caption{Comparison between Goodness of Fit for various $T_{eff}$ correlations.}
\label{tab:Goodness of fit for Teff}
\resizebox{0.75\textwidth}{!}{
\begin{tabular}{cccccccccc} 
   \hline
   \hline
Index & T & N & R & Rsqr & SEE & m0$\dagger$ & a$\dagger$ & Relation$\dagger$ & Remarks* \\	
\hline

Case 1: & & & & & & & & &\\
$^{12}$CO at 2.29 $\mu$m (CO229) & 101 & 84 & 0.97 & 0.93 & 130 & 5651 $\pm$ 44 & --99 $\pm$ 03 & Eq.~\ref{equation:linear_individuallines} & (--0.3, +0.3) \\
$^{12}$CO at 2.32 $\mu$m  (CO232) & 107 & 85 & 0.96 & 0.92 & 150 & 5794 $\pm$ 55 & --144 $\pm$ 05 & Eq.~\ref{equation:linear_individuallines} & (--0.3, +0.3) \\
& & & & & & & & &\\
Case 2: & & & & & & & & &\\
CO229   & 106 & 96 & 0.96 & 0.92 & 124 & 5486 $\pm$ 46 & --87 $\pm$ 03 & Eq.~\ref{equation:linear_individuallines} & (0.0, +0.96) \\
CO232   & 106 & 97 & 0.94 & 0.88 & 146 & 5502 $\pm$ 57 & --117 $\pm$ 04 & Eq.~\ref{equation:linear_individuallines} & (0.0, +0.96) \\
& & & & & & & & & \\
Case 3: & & & & & & & & & \\
CO229   & 158 & 131 & 0.93 & 0.87 & 180 & 5290 $\pm$ 41 & --80 $\pm$ 03 & Eq.~\ref{equation:linear_individuallines} & (--1.81, 0.0) \\
CO232   & 158 & 137 & 0.92 & 0.84 & 204 & 5375 $\pm$ 49 & --113 $\pm$ 04 & Eq.~\ref{equation:linear_individuallines} & (--1.81, 0.0) \\
& & & & & & & & & \\
Case 4: & & & & & & & & & \\
CO229   & 260 & 218 & 0.95 & 0.90 & 149 & 5370 $\pm$ 30 & --82.7 $\pm$ 02 & Eq.~\ref{equation:linear_individuallines} & (--1.77, 0.96) \\
CO232   & 260 & 232 & 0.92 & 0.85 & 186 & 5398 $\pm$ 38 & --113 $\pm$ 03 & Eq.~\ref{equation:linear_individuallines} & (--1.81, 0.96) \\

\hline
\end{tabular}}

T $-$ total nos. of data points; N $-$ no. of points used for fitting after 2$\sigma$ clipping \\ 
R $-$ correlation coefficient; Rsqr $-$ coefficient of determination;
SEE $-$ standard error of estimate \\ 
$\dagger$ $-$ Relation (equation) used to establish the correlation, m0 and a are coefficients of the equation\\ * $-$ Metallicity range of the stars after 2$\sigma$ clipping
  \\
\end{table*}

\subsubsection{Metallicity indicator} \label{Metallicity_correlation}
All four lines (NaI, FeI, CO229, and CO232) are undertaken to study the metallicity dependence of those lines and to establish the new empirical relations. Three different cases are exercised to establish empirical relations for a simple and accurate estimation of [$Fe/H$] as follows.

\begin{figure*}
	\includegraphics[scale=0.55]{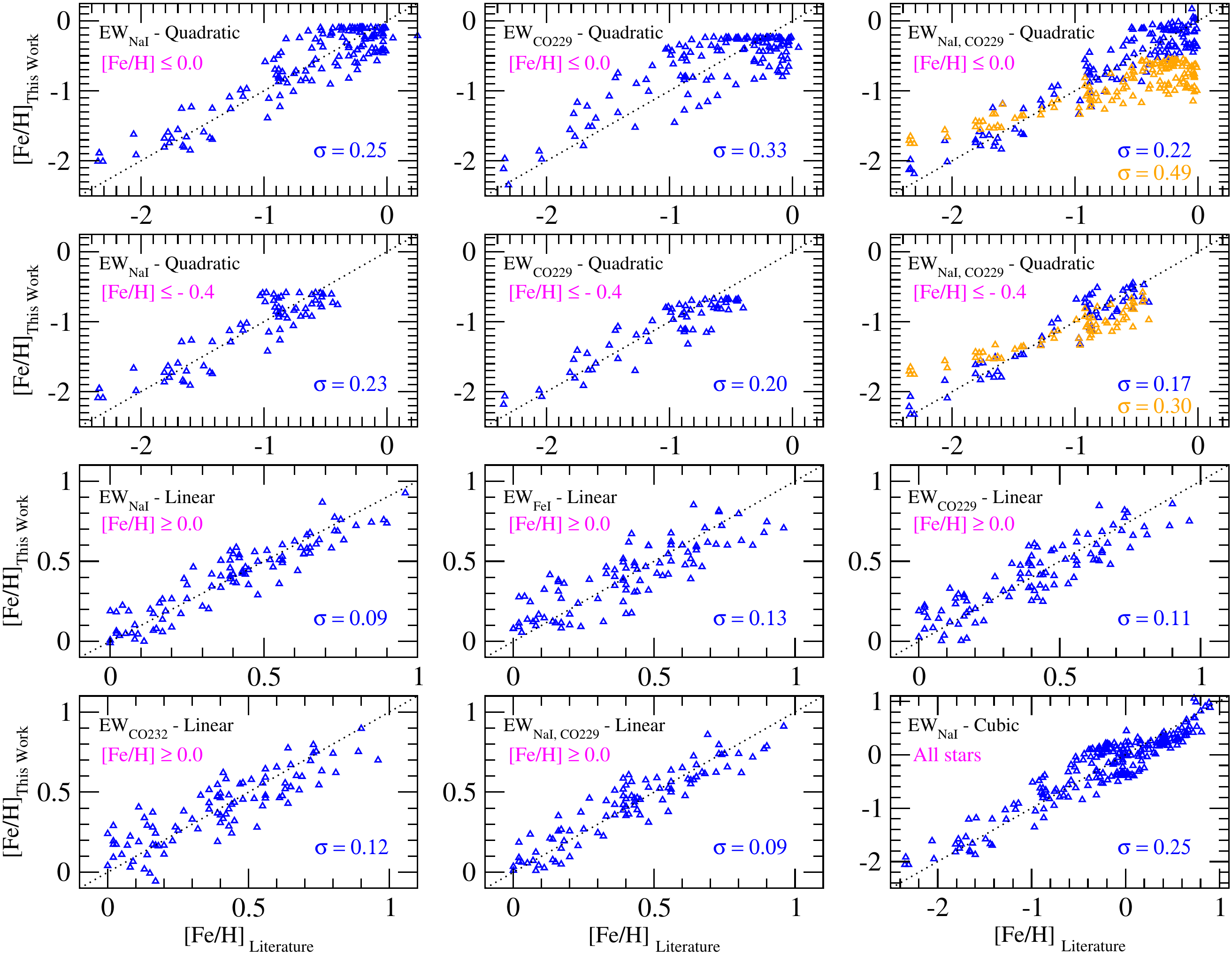}
   \caption{[$Fe/H$] comparison between derived values from the empirical relations established in this work and literature values (blue triangles). The orange triangles show the estimation using \citet{2001AJ..122..1896} quadratic empirical relation after excluding the coefficient of Ca I at 2.26 $\mu$m line discussed in the text.  The dotted line displays the one-to-one correspondence of metallicity.}
    \label{Fig:index-meta_correlation_residual}
\end{figure*}

First (Case 1m), we consider those stars having [$Fe/H$] $\leq$ 0.0 dex. We first explore a simple linear (using Eq.~\ref{equation:linear_individuallines} and Eq.~\ref{equation:linear_combinedlines}) fit and find the large SEE ($>$ 0.35 dex) for individual lines as well as for the combination of lines. Therefore, we carry out the quadratic fits to establish the empirical relations between metallicity and indices. For quadratic fits, we follow equations
\begin{equation} \label{equation:quadratic_individuallines}
z = m0 + a \times x +  c \times x^2 
\end{equation}
for a individual line, and
\begin{equation} \label{equation:quadratic_combinedlines}
z = m0 + a \times x +  b \times y + c \times x^2 + d \times y^2 
\end{equation}
for a combination of two lines, where z = [Fe/H], x and y are EWs of spectral features, and m0, a, b, c and d are the coefficients of the fit. The quadratic fit (Eq.~\ref{equation:quadratic_individuallines}) of NaI and CO229 yield the metallicity scale with a typical accuracy of 0.25 dex and 0.33 dex, respectively. However, a quadratic fit of FeI and CO232 lines provides the larger SEE (i.e. SEE $>$ 0.33) and henceforth, those lines are ignored. We then investigate a combination of lines (using Eq.~\ref{equation:quadratic_combinedlines}) and find that the best empirical relation is provided by a quadratic fit of [$Fe/H$] to the NaI and CO229 spectroscopic indices. The typical accuracy of [$Fe/H$] estimation is 0.22 dex. The parameters of the fit are listed in Table~\ref{tab:Goodness of fit}. We then invert the process and calculate [$Fe/H$] of the sample stars using the above established quadratic relation. We further evaluate [Fe/H] using the quadratic equation of \citet{2001AJ..122..1896}. The comparison of both measurements with the literature value is illustrated in Fig.~\ref{Fig:index-meta_correlation_residual}, where the blue triangles refer to the estimation from our empirical relation and orange triangles represent the measurement using the empirical relation of \citet{2001AJ..122..1896}. \citet{2001AJ..122..1896} established empirical relations using EWs of NaI, CaI, and CO229 of 105 stars. However, we here exclude the coefficient of CaI because of the problem of X-Shooter spectra as mentioned earlier. The mean and standard deviation of the fit residuals are $\Delta [Fe/H]_{Avg}$ = 0.01 dex, $\sigma_{[Fe/H]}$ = 0.22 dex for our empirical relation and $\Delta [Fe/H]_{Avg}$ = 0.32 dex $\sigma_{[Fe/H]}$ = 0.49 dex for \citet{2001AJ..122..1896} empirical relation with respect to the literature value. It is evident that our measurements of [$Fe/H$] are not in good agreement with the measured values from the empirical relation of \citet{2001AJ..122..1896}; in fact, \citet{2001AJ..122..1896} empirical relation based estimation overestimates [Fe/H] below --1.2 dex and underestimates above --1.2 dex. To investigate whether this discrepancy is due to the exclusion of CaI line from the empirical relation of \citet{2001AJ..122..1896} or not, we consider that the EWs of CaI are alike to EWs of NaI and redo our calculation. However, our results do not change significantly which indicates that the difference does not arise because of the exclusion of CaI from the empirical relation. This investigation also confirms the relative lack of sensitivity of the empirical relation to the CaI line which was already seen by \citet{2001AJ..122..1896}. Therefore, the possible reasons for the discrepancy are the use of a different sample of stars for the empirical relations and the accuracy of the [$Fe/H$] estimation of those stars used for calibration.

We then narrow down the metallicity range by considering only those stars having [$Fe/H$] $\leq$ --0.4 dex as the spectral lines begin to saturate above that metallicity and the sensitivity of those lines to [Fe/H] appear to decrease (see, Fig.~\ref{Fig:index-meta_behaviour}). We find better empirical relations and the SEE of those relations are significantly improved as shown in Table~\ref{tab:Goodness of fit} and Fig.~\ref{Fig:index-meta_correlation_residual}. Furthermore, we again measure [$Fe/H$] using the empirical relation of \citet{2001AJ..122..1896} in this narrow range. Although the standard deviation is significantly improved, it ($\sigma_{[Fe/H]}$ $\sim$ 0.3 dex) still is not in agreement with our established empirical relation based measurement ($\sigma_{[Fe/H]}$ $\sim$ 0.17 dex). Note that the spectral lines become very weak below --1.8 dex (see, Fig.~\ref{Fig:index-meta_behaviour}). Therefore, our relations need to be considered with care below this metallicity.

Second (Case 2m), the sample giants having [Fe/H] $\geq$ 0.0 dex are undertaken for the empirical relation. We first explore linear fits (using Eq.~\ref{equation:linear_individuallines}) for all individual spectral lines after excluding the limiting 2$\sigma$ outliers. The parameters of the fit are listed in Table~\ref{tab:Goodness of fit}. We find that all the individual lines are a good metallicity indicator in this range, however, NaI line with a typical accuracy of 0.09 dex yields the best empirical relation. We also explore all possible linear and quadratic combinations of multi-lines. However, those relations do not improve the accuracy of the best correlation (SEE $\sim$ 0.09 dex). For example, the linear fitting parameters of NaI and CO229 combination lines are listed in Table~\ref{tab:Goodness of fit}. We then invert the process and calculate [Fe/H] of the fitted sample stars using all empirical relations as depicted in Fig.~\ref{Fig:index-meta_correlation_residual}.

In the third and final case (Case 3m), all the sample is considered for the empirical relation. Here, we examine only NaI line because --- first, the NaI carries more weight in our multi-lines relations in spite of the fact that EWs of CO229 is many times stronger than the former, and second, it is less sensitive to $T_{eff}$ than CO lines. We apply a cubic equation to find the best fit for the data as
\begin{equation} \label{equation:cubic_individuallines_meta}
z = m0 + a \times x +  c \times x^2 + e \times x^3,
\end{equation}
 where z = [Fe/H], x = NaI, and m0, a, c and e are the coefficients of the fit. The fit parameters are presented in Table~\ref{tab:Goodness of fit}. The SEE of an estimate of [$Fe/H$] from this fit is 0.25 dex, while the value of R is 0.93. Although the SEE of the relation is greater than the SEE of Case 1m and Case 2m, this scale can offer an initial [$Fe/H$] measurement. The relation is also advantageous for cluster stars because of a considerably less relative star to star scatter within a cluster for the NaI line than CO lines \citep{2001AJ..122..1896}. Similar to the previous cases, Fig.~\ref{Fig:index-meta_correlation_residual} illustrates the inverse process and shows the comparison between our measurements from the empirical relation and the literature value.

We develop a reliable, accurate technique based on near-IR spectroscopy which can apply for measuring the
metallicity in the range --1.80 dex to +0.96 dex. The differences in the metal-poor regime to other studies are because of the different sample of stars used for the correlation and different methods adopted  to estimate parameters of sample stars. However, care should be taken to measure metallicity for metal-rich stars. The majority of our sample stars in the metal-rich regime are taken from \citet{2015ApJ...809..143D}, and their measurements uncertainties may be underestimated, especially for metal-rich stars, because of the systematics in the model \citep{2015ApJ...809..143D}.


 \begin{table*}
\centering
\caption{Comparison between Goodness of Fit for various correlations.}
\label{tab:Goodness of fit}
\resizebox{0.94\textwidth}{!}{
\begin{tabular}{lcccccccccccc} 
   \hline
   \hline
Index & T & N & R & R$_{sqr}$& SEE & m0$\dagger$ & a$\dagger$ & b$\dagger$ & c$\dagger$ & d$\dagger$ & e$\dagger$ & Relation$\dagger$  \\	
\hline
\textbf{[Fe/H] $\leq$ 0.0} & & & & & & & & & & & & \\
 & & & & & & & & & & & & \\
 x=EW$_{NaI}$                 & 154 & 142 & 0.91 & 0.83 & 0.25 & $-$2.237 $\pm$ 0.070 & 1.551 $\pm$ 0.080 & ... & $-$0.280 $\pm$ 0.020 & ... & ... & Eq.~\ref{equation:quadratic_individuallines} \\  
 x=EW$_{CO229}$               & 154 & 134 & 0.82 & 0.67 & 0.33 & $-$2.342 $\pm$ 0.126 & 0.255 $\pm$ 0.021 & ... & $-$0.006 $\pm$ 0.001 & ... & ... & Eq.~\ref{equation:quadratic_individuallines} \\
 x=EW$_{NaI}$, y=EW$_{CO229}$ & 154 & 135 & 0.94 & 0.88 & 0.22 & $-$2.394 $\pm$ 0.073 & 1.423 $\pm$ 0.101 & 0.055 $\pm$ 0.017 & $-$0.218 $\pm$ 0.023 & $-$0.003 $\pm$ 0.001 & ... & Eq.~\ref{equation:quadratic_combinedlines} \\
 
  & & & & & & & & & & & &  \\
 \textbf{[Fe/H] $\leq$ $-$0.40} & & & & & & & & & & & & \\
 & & & & & & & & & & & & \\
 x=EW$_{NaI}$                 & 82 & 70 & 0.91 & 0.82 & 0.23 & $-$2.348 $\pm$ 0.086 & 1.779 $\pm$ 0.157 & ... & $-$0.449 $\pm$ 0.062 & ... & ... & Eq.~\ref{equation:quadratic_individuallines}\\  
 x=EW$_{CO229}$               & 82 & 64 & 0.92 & 0.85 & 0.20 & $-$2.370 $\pm$ 0.100 & 0.184 $\pm$ 0.017 & ... & $-$0.005 $\pm$ 0.001 & ... & ... & Eq.~\ref{equation:quadratic_individuallines}\\
 x=EW$_{NaI}$, y=EW$_{CO229}$ & 82 & 64 & 0.95 & 0.91 & 0.17 & $-$2.478 $\pm$ 0.070 & 1.044 $\pm$ 0.133 & 0.108 $\pm$ 0.018 & $-$0.187 $\pm$ 0.043 & $-$0.004 $\pm$ 0.001 & ... & Eq.~\ref{equation:quadratic_combinedlines} \\
 
  & & & & & & & & & & & & \\
\textbf{[Fe/H] $\geq$ $-$0.0} & & & & & & & & & & & & \\
 & & & & & & & & & & &  \\
 x=EW$_{NaI}$                 & 106 & 94 & 0.93 & 0.86 & 0.09 & $-$0.226 $\pm$ 0.028 & 0.168 $\pm$ 0.007 & ... & ... & ... & ... & Eq.~\ref{equation:linear_individuallines}\\
 x=EW$_{FeI}$                 & 106 & 92 & 0.85 & 0.71 & 0.13 & $-$0.042 $\pm$ 0.030 & 0.366 $\pm$ 0.024 & ... & ... & ... & ... & Eq.~\ref{equation:linear_individuallines}\\ 
 x=EW$_{CO229}$               & 106 & 96 & 0.89 & 0.79 & 0.11 & $-$0.345 $\pm$ 0.042 & 0.047 $\pm$ 0.003 & ... & ... & ... & ... & Eq.~\ref{equation:linear_individuallines} \\ 
 x=EW$_{CO232}$               & 106 & 94 & 0.86 & 0.74 & 0.12 & $-$0.359 $\pm$ 0.049 & 0.063 $\pm$ 0.004 & ... & ... & ... & ... & Eq.~\ref{equation:linear_individuallines}\\       
 
 x=EW$_{NaI}$, y=EW$_{CO229}$ & 106 & 93 & 0.93 & 0.86 & 0.09 & $-$0.262 $\pm$ 0.036 & 0.134 $\pm$ 0.018 & 0.011 $\pm$ 0.005 & ... & ... & ... & Eq.~\ref{equation:linear_combinedlines}\\
   & & & & & & & & & & & & \\
Case 3m: & & & & & & & & & & & &  \\
\textbf{All stars} & & & & & & & & & & & &  \\
 & & & & & & & & & & & & \\
 x=EW$_{NaI}$ & 260 & 227 & 0.93 & 0.87 & 0.25 & $-$2.328 $\pm$ 0.080 & 1.863 $\pm$ 0.109 & ... & $-$0.485 $\pm$ 0.042 & ... & 0.045 $\pm$ 0.005 & Eq.~\ref{equation:cubic_individuallines_meta} \\

\hline
\end{tabular}} 

T - total nos. of data points; N - no. of points used for fitting after eliminating 2$\sigma$ outliers \\ 
R - correlation coefficient; Rsqr $-$ coefficient of determination; SEE - standard error of estimate \\ 
$\dagger$ $-$ Relation (equation) used to establish the correlation, m0, a, b, c, d and e are coefficients of the equation\\ 
\end{table*} 

\section{Systematic Error Sources}\label{Systematic_errors}
In this section, we investigate the various sources of systematics that can impact our results. Systematic errors can arise in the EWs measurement between data from the two instruments. Different resolutions of the two instruments and the presence of sky emission lines or telluric absorption lines near the feature and/or continuum bands used for EW estimation can cause systematic errors. However, we degrade the resolution of both data to the same resolution (R $\sim$ 1200) before estimating EW and showed that there is no significant resolution effect of EW estimation (see Section~\ref{Data_collection}). Also, no sky emission line or telluric absorption line is evident in the wavelength region of our interest in EW estimation. Therefore, we can rule out systematics in EWs measurement that can impact our results.

Since different measurement techniques can lead to large discrepancies in the parameters and the abundances of the same stars (e.g., see \citealt{2016ApJS..226....4H, 2019MNRAS.486.2075B}), systematic errors can be expected in parameter estimation between the NIFS and X-shooter data sets since they are based on two diverse measurement techniques. One possible way to check these systematics is to compare the estimated parameters of common stars in both methods. However, no common star is present between the two data sets. We used the data itself to estimate any possible systematics between the two data sets via Bayesian inference. 
We choose X-shooter stellar parameters estimated by \citet{2019A&A...627A.138A} as our model's predicted stellar parameters and added an offset to NIFS stellar parameters (estimated by \citealt{2015ApJ...809..143D}). This offset was supported by an informative prior in our model. The informative prior for the systematic offsets in the stellar parameters were modelled as a normal distribution with a mean and a standard deviation inferred from a chain of different stellar parameters studies connecting the X-shooter estimation to the NIFS estimation. Additional details of Bayesian fit and subsequent analysis are presented in Appendix \ref{Bayesian_model}. Using the T$_{eff}$--CO229 fit, we derived the offset for T$_{eff}$ in NIFS estimation to be 42 $\pm$65 K. Similarly, we estimated systematic in [Fe/H] from NaI--[Fe/H] linear fit to be --0.15 $\pm$0.09. 
This analysis shows that there is no large systematics in parameter estimation by two different methods and confirms the fact that the effect of metallicity on $T_{eff}$--CO empirical relations discussed in Section~\ref{Effective_temperature_correlation} are not simply the result of systematic differences between the two data sets. We also confirm this effect from Bayesian analysis considering stars with [$Fe/H$] $> +0.3$ dex and [$Fe/H$] $< -0.3$ dex as illustrated in Section \ref{MetaEffectOnTeffCO229}. Since the Bayesian model allows us to incorporate uncertainties in systematics self consistently in the inference, we have done a parallel analysis of the stellar parameter versus EW relations in Appendix \ref{Bayesian_model}.


\section{Summary and Conclusions} \label{Summary_and_Conclusions}

In this paper, we make use of 260 cool giants having a wider metallicity coverage than in earlier work to present a method, based on low-resolution NIR $K$-band spectroscopy (R $\sim$ 1200) of individual stars, for the precise estimation of fundamental parameters for the cool giants. We measure equivalent widths of some of the prominent $K$-band spectral features like Na I at 2.20 $\mu$m, Fe I at 2.23 $\mu$m and  $^{12}$CO at 2.29$\mu$m and 2.32 $\mu$m. We have investigated the behavior of those EWs with fundamental parameters (e.g. effective temperature and metallicity).  The main results in this work can be summarized as follows.

\begin{enumerate}

\item We establish new empirical relations between effective temperature and equivalent widths $^{12}$CO at 2.29 $\mu$m and 2.32 $\mu$m. We confirm that $^{12}$CO at 2.29 $\mu$m is a very good indicator of effective temperature. We show a detailed quantitative metallicity dependence of effective temperature -- CO empirical relations considering the stars of four different metallicity ranges and we find that the empirical relations based on solar-neighborhood stars can incorporate large uncertainty in evaluating $T_{eff}$ for more metal-poor or metal-rich stars. We also find no significant effect of equivalent widths estimation on resolution degradation from R$\sim$ 5400 to R$\sim$ 1200. Thus, effective temperature -- CO empirical relations could be used more generally.

\item We obtain new empirical relations between metallicity and the spectral features for metal-rich and metal-poor stars. We show that the quadratic fit of the combination of Na I and $^{12}$CO at 2.29 $\mu$m lines is an excellent metallicity indicator at [$Fe/H$] $\leq$ --0.4 dex, whereas a linear empirical relation of any lines studied here yields metallicity with good accuracy at [$Fe/H$] $\geq$ 0.0 dex.

\end{enumerate}  

We expect that this work will help for precise estimation of the effective temperature and the metallicity of stars using the NIR spectral region and to exploit in-depth the so far poorly-studied heavily obscured regions. Our new diagnostic tools are very easy to use and need not require knowledge of the reddening and distance to the object.

\section*{Acknowledgements}
The authors are very much thankful to the reviewer for his/her critical and valuable comments, which helped us to improve the paper. This research work is supported by the Tata Institute of Fundamental Research, Mumbai under the Department of Atomic Energy, Government of India. SG is thankful to Dr. T. Do for sharing the NIFS data. SG and DKO acknowledge the support of the Department of Atomic Energy, Government of India, under project Identification No. RTI 4002. This research has made use of the SIMBAD database, operated at CDS, Strasbourg, France.

\section*{DATA AVAILABILITY}
All observational data utilized in this paper are publicly available and can be found at: \url{http://xsl.astro.unistra.fr/} (X-shooter data) and \url{https://zenodo.org/record/3606913/} (NIFS data). Table~\ref{tab:all_measured_ews} is available in its entirety as online supplementary material.




\bibliographystyle{mnras}
\bibliography{supriyo_reference-v1} 





\appendix

\section{Bayesian fit of EWs and stellar parameters} \label{Bayesian_model}
In order to cross-check the statistical validity of the derived empirical relationships, we carried out an independent analysis of the equivalent widths (EWs) versus stellar parameters in the Bayesian framework. The Bayesian framework enables us to model systematic bias between the X-shooter and NIFS data, and propagate forward all the uncertainties in a self-consistent manner to the final empirical relationships.
\subsection{Systematics between X-shooter and NIFS stellar parameters} \label{SystematicsModelling}
\begin{figure}
	\includegraphics[width=0.49\textwidth,angle=0]{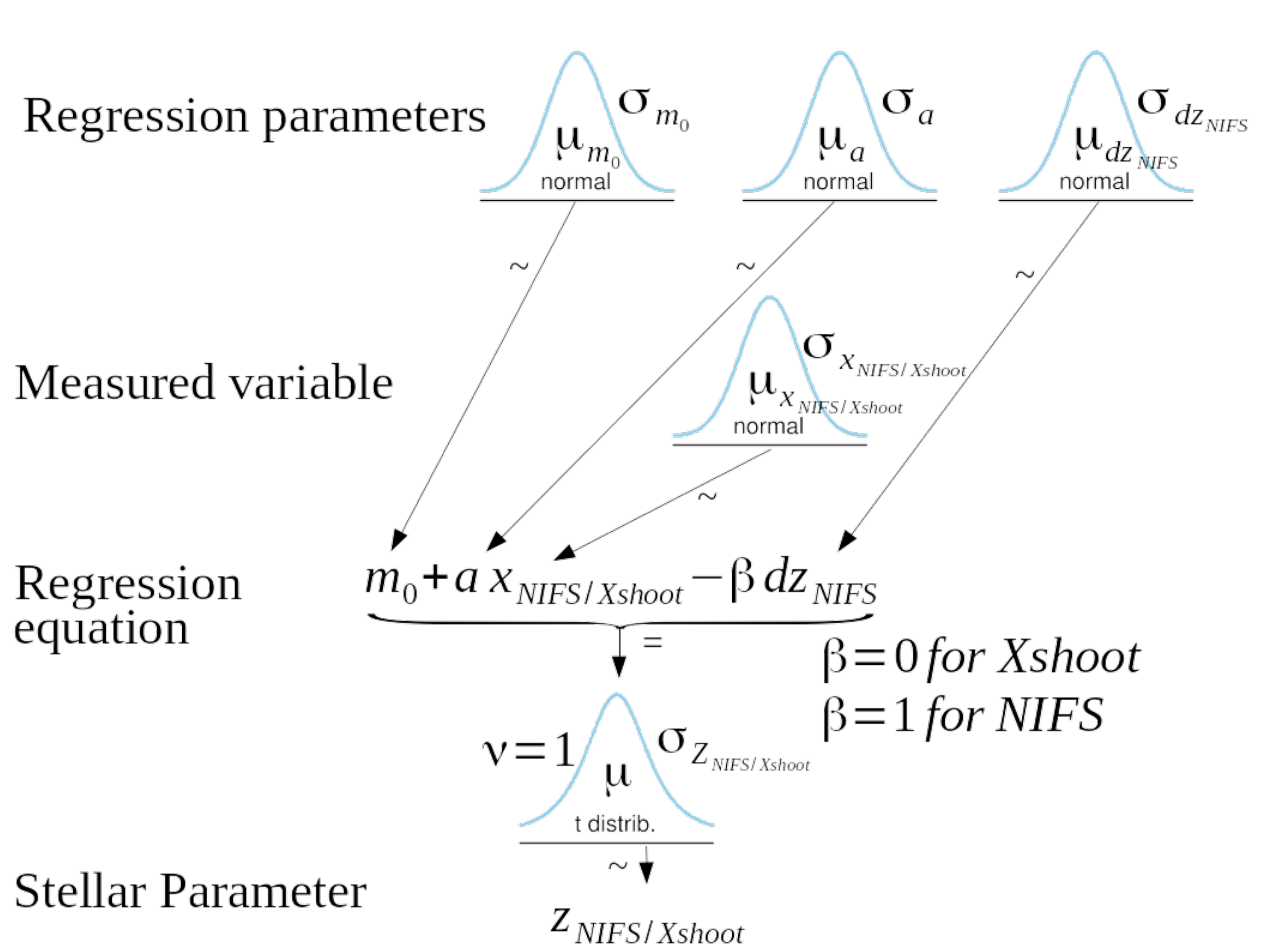}
    \caption{Kruschke style diagram of the Bayesian model of the regression formula \ref{equation:bayesian_linear_SPversusEW}. This diagram shows the probabilistic distribution from which each of the variable in the regression formula is shown here. During the Monte-Carlo run to fit the Bayesian model, each of the variables in the iteration will be sampled for these distributions. One can also read this diagram as a representation of the forward model to obtain the stellar parameter starting with EWs.}
    \label{Fig:KruschkeDiagramLinModel}
\end{figure}

As discussed in Section \ref{Systematic_errors}, due to different techniques used in the estimation of the X-shooter stellar parameters by \citet{2019A&A...627A.138A} and NIFS stellar parameters by \citet{2015ApJ...809..143D}, there could be systematics between the two data sets. In the absence of any common stars between the two data sets, we use the scatter in the data itself to model the systematics between the two data-sets. We use a subset of data points from both the studies where they overlap in stellar parameter versus EW parameter space to constrain the systematics between them. There is a significant linear trend in the data inside this region. Hence, we use the probabilistic reformulation of the linear equation \ref{equation:linear_individuallines} to model this data,
\begin{equation} \label{equation:bayesian_linear_SPversusEW}
\begin{aligned}
z_{X-shooter/NIFS} \sim {} & \mathcal{N}(m_0 + a \times x_{X-shooter/NIFS} \\
                           & -\beta dz_{NIFS},\,\sigma^{2}_{z_{X-shooter/NIFS}})\,,
\end{aligned}
\end{equation}

where $x_{i} \sim \mathcal{N}(EW_i, \sigma^{2}_{EW_i})\,$ and binary variable $\beta$ is 0 for X-shooter data, and 1 for NIFS data. Similar to Equation \ref{equation:linear_individuallines} here $z$ is fundamental stellar parameter (e.g., T$_{eff}$) for the NIFS and X-shooter data, $x$ is the EWs of spectral features, $m_0$, $a$ is the coefficients of the slope fit.  This formalism is visually represented by Kruschke style diagram in Figure \ref{Fig:KruschkeDiagramLinModel}. In the model we consider the stellar parameter $z$ (e.g., T$_{eff}$) to be drawn from a Student T distribution to make the inference robust against outliers due to stellar contamination. The measured EW quantities ($x$) are drawn from the normal distribution defined by the measured mean and sigma. Thus the Bayesian formalism naturally enables us to incorporate both the measurement error in the predictor variable $x$ and the model error in stellar parameter $z$ in the regression problem. The extra term $dz_{NIFS}$ is the systematic offset between the X-shooter stellar parameter and NIFS parameter, which we will infer self consistently from the data.  The newly added terms  $dz_{NIFS}$ is partly degenerate with $m_0$ term.  Therefore, the prize we are paying for an unbiased estimate of the $m_0$ term is its larger variance. By treating $dz_{NIFS}$ as a correction term only for NIFS stellar parameter (using the $\beta$ flag), we are implicitly adopting the X-shooter stellar parameter estimates as our model's predicted stellar parameter. A major assumption in this modelling is that the systematic $dz_{NIFS}$ is a constant number across the entire Metallicity space or Temperature space in this study.

\subsubsection{Informative Prior for $dz_{NIFS}$}
As there are no common stars between the two data sets, we compared different literature to find conservative systematic offsets between X-shooter and NIFS parameters estimation as listed in Table~\ref{tab:Estimation_systematic_offsets}. 

\begin{table}
\centering
\caption{Literature comparison to estimate systematic offsets between X-shooter and NIFS data sets.}
\label{tab:Estimation_systematic_offsets}
\resizebox{0.45\textwidth}{!}{
\begin{tabular}{llll} 
   \hline
Parameter & L1 -- L2 & L2 -- L3 & L4 -- L3 \\	 
\hline		
$\Delta$ T$_{eff}$ (K)  & $\mu$= --15, $\sigma$= 70     & $\mu$=--48, $\sigma$= 94    &$\mu$=50, $\sigma$= 400 \\
$\Delta$ [$Fe/H$] (dex) & $\mu$= --0.02, $\sigma$= 0.09 & $\mu$= 0.04, $\sigma$= 0.07 &$\mu$=--0.2, $\sigma$= 0.3 \\
\hline
\end{tabular}
} 
\\
Ref : L1 -- \citet{2019A&A...627A.138A}, L2 -- \citet{2011A&A...525A..71W},\\ L3 -- \citet{2013aap...549A.129}, L4 -- \citet{2015ApJ...809..143D}. \\ $\Delta$ T$_{eff}$ = residual of T$_{eff}$, $\Delta$ [$Fe/H$] = residual of [$Fe/H$] \\
$\mu$ = mean, $\sigma$ = standard deviation
\end{table}

Based on this comparison, we chose an informative prior for the systematic offsets in NIFS data (\citealt{2015ApJ...809..143D}) to match X-shooter data (\citealt{2019A&A...627A.138A}) as normal distributions with the following mean ($\mu$) and standard deviation: Teff$_{NIFS}$ = -113 K $\pm$ 416 K and [Fe/H]$_{NIFS}$ = 0.22 dex $\pm$ 0.32 dex.

\subsubsection{Posterior for $dz_{NIFS}$}
\begin{figure}
	\includegraphics[width=0.5\textwidth,angle=0]{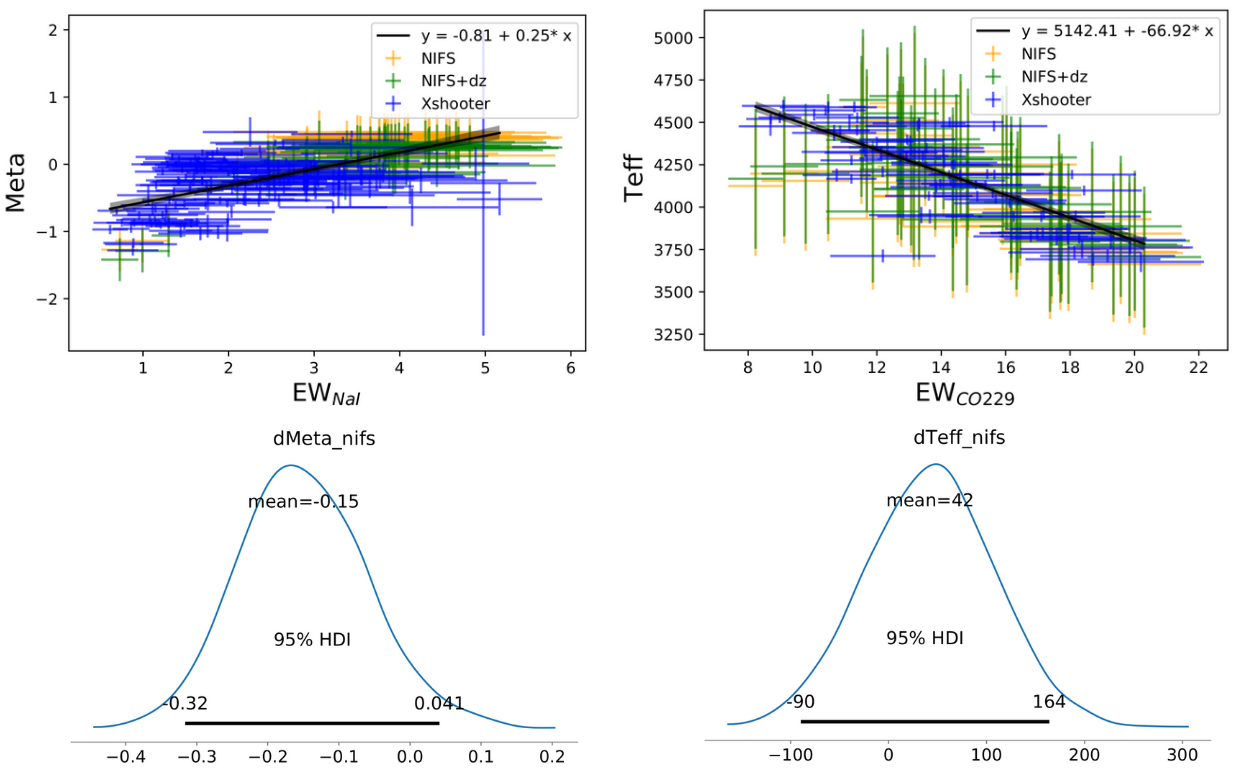}
    \caption{Bayesian fit of a linear model in the overlapping parameter space of the NIFS and X-shooter data is shown in the top panel. Black curve is the best fit, and the grey shaded region is the 1 sigma interval of the fitted model. The blue points are the X-shooter data, yellow points are the NIFS data, and green points are the NIFS data offset by the best estimate of the stellar parameter systematic $dz_{NIFS}$. The posterior distribution of $dz_{NIFS}$ corresponding to each stellar parameter fit is shown in the bottom panel. The 95\% Highest Density Interval (HDI) is also marked inside the posterior distributions.}
    \label{Fig:dz_fits}
\end{figure}

We implemented our Bayesian model in PyMC3 \citep{10.7717/peerj-cs.55}. PyMC3 uses a No-U-Turn Sampler (NUTS), a self-tuning variant of Hamiltonian Monte Carlo (HMC) to fit the model. We discarded the first 500 points for burn in and sampled another 2000 points. Three independent chains were run and they all converged to the same posterior distribution. Figure \ref{Fig:dz_fits} shows the poster obtained for the systematic term $dz_{NIFS}$ from the different combinations of the stellar parameter versus EW fits. Table \ref{tab:dz_posterior} summarises the mean and sigma we adopt as our posterior from this analysis for $dz_{NIFS}$, as well as our highly informative prior for the $dz_{NIFS}$ in all of our further analysis.

\begin{table}
\centering
\caption{Estimate of the systematic offsets between X-shooter and NIFS data sets.}
\label{tab:dz_posterior}
\resizebox{0.45\textwidth}{!}{
\begin{tabular}{cl} 
   \hline
Parameter $z$ & Posterior for $dz_{NIFS}$ \\	 
\hline		
T$_{eff}$ (K)  & $\mu$= 42, $\sigma$= 65 \\
$[Fe/H]$ (dex) & $\mu$= --0.15, $\sigma$= 0.09 \\
\hline
\end{tabular}
} 
\end{table}

\subsection{Best model for [$Fe/H$] versus EW$_{NaI}$} \label{ModelSelectionMeta_NaI}
\begin{figure}
	\includegraphics[width=0.5\textwidth,angle=0]{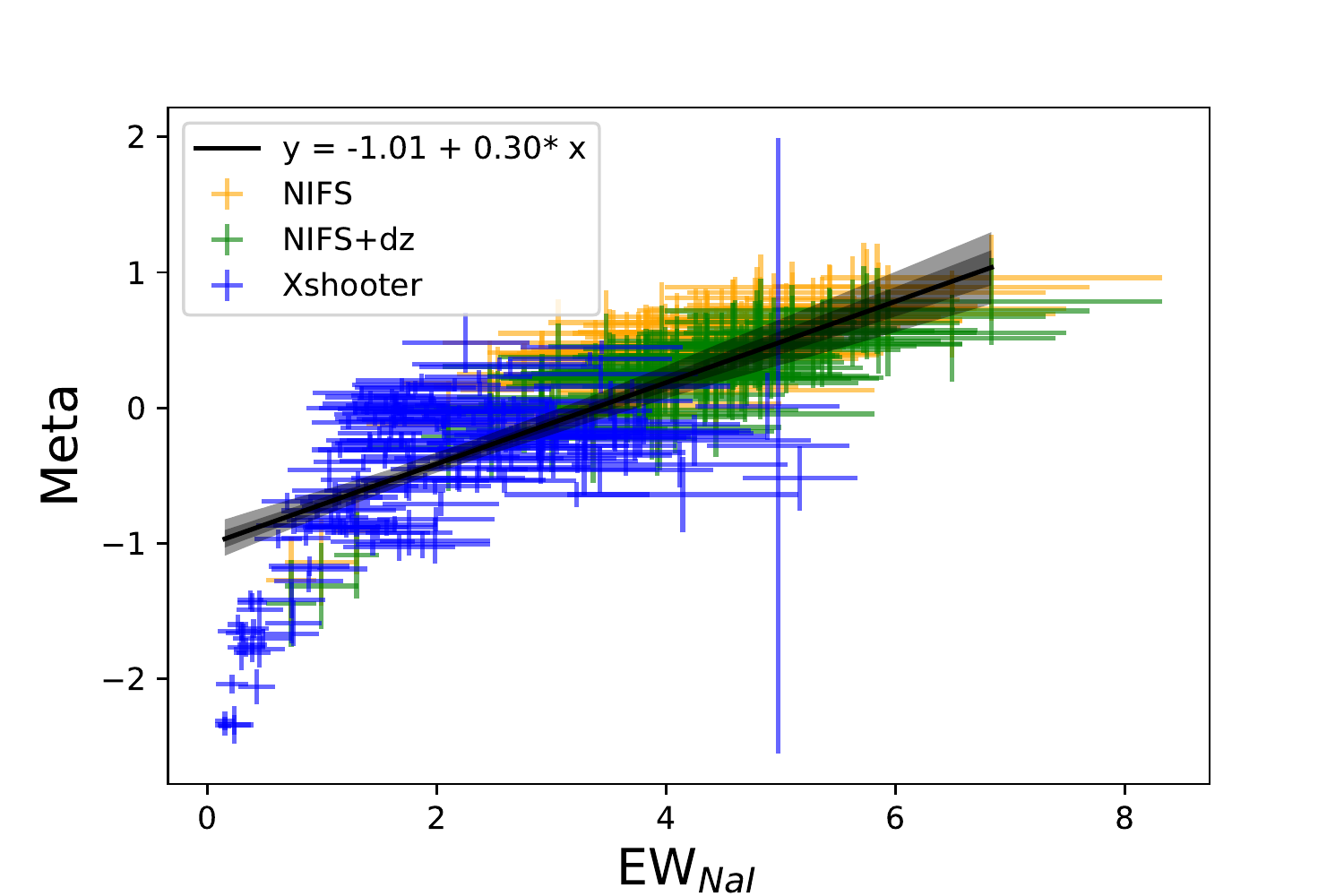}
	\includegraphics[width=0.48\textwidth,angle=0]{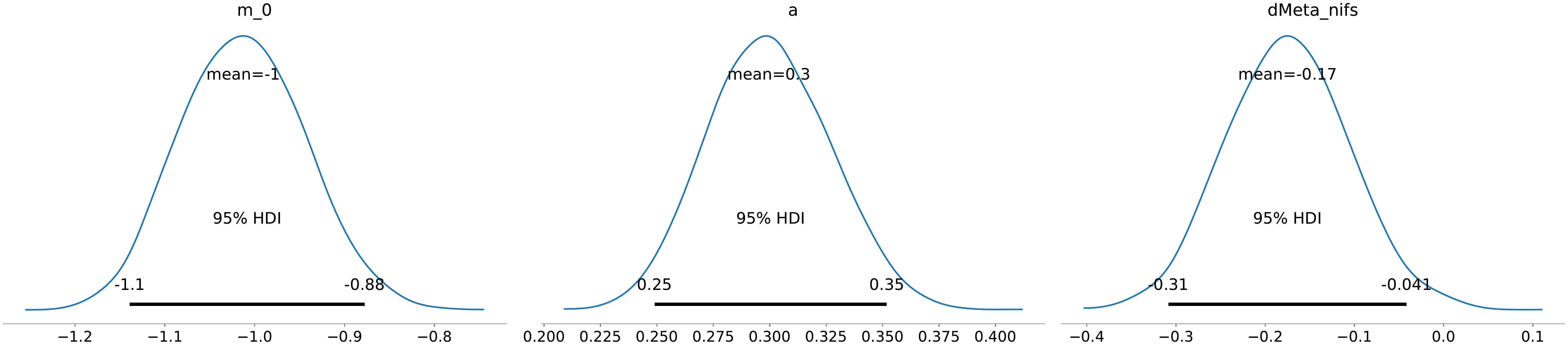} 
    \caption{Bayesian fit of the linear model in [$Fe/H$] versus EW$_{NaI}$ is shown. Black curve is the best fit, and the grey shaded regions are  the 1 and 2 sigma intervals of the fitted model. The blue points are the X-shooter data, yellow points are the NIFS data, and green points are the NIFS data offset by the best estimate of the stellar parameter systematic $dMeta_{NIFS}$. The bottom panel displays the posterior distributions of the coefficients in the model. The 95\% Highest Density Interval (HDI) is also marked inside the posterior distributions. The summary of the posterior distributions and BIC are tabulated in Table \ref{tab:MetaEWNaI_fits}.}
    \label{Fig:MetaEWNaI_fits_linear}
\end{figure}

\begin{figure}
	\includegraphics[width=0.48\textwidth,angle=0]{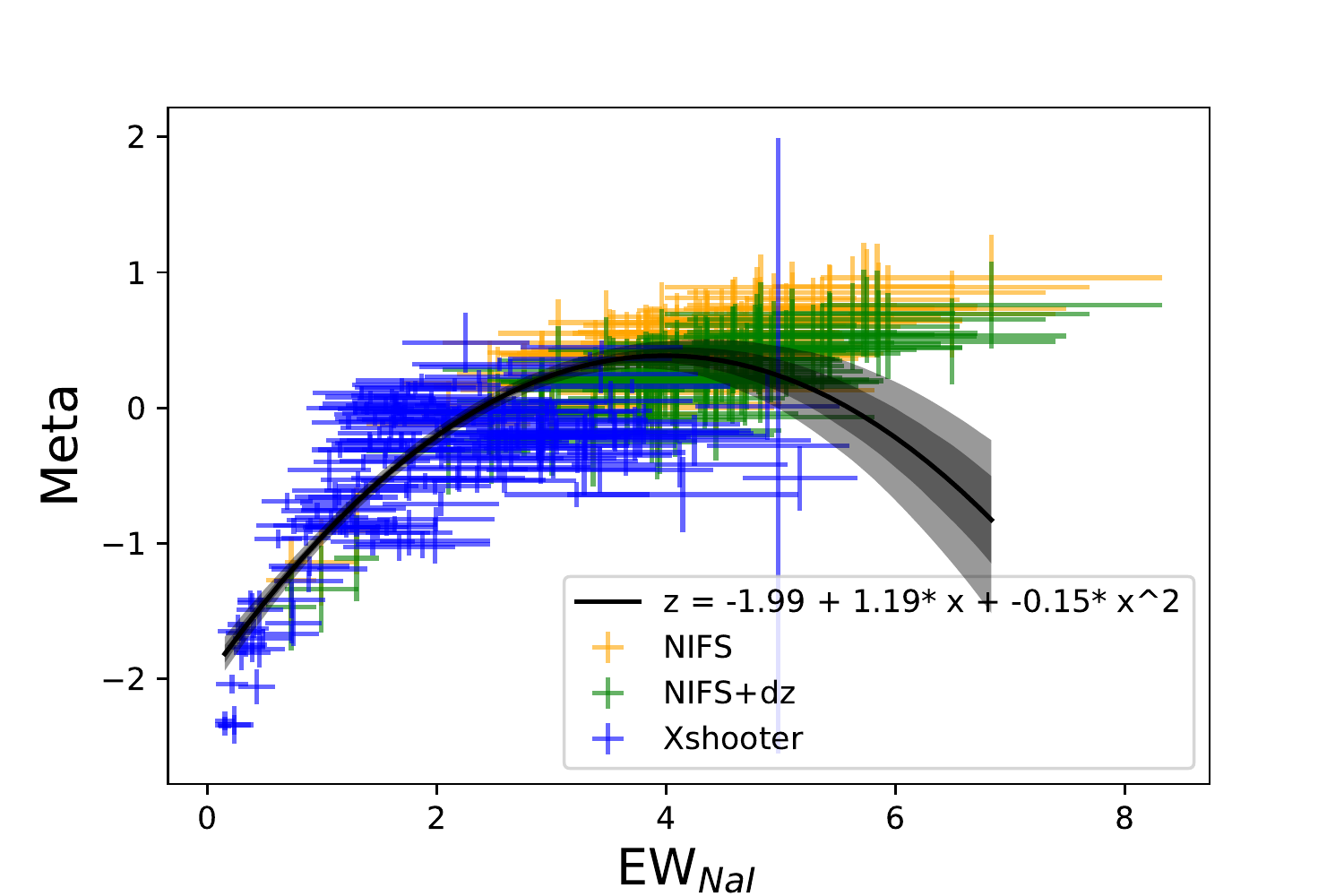} 
	\includegraphics[width=0.48\textwidth,angle=0]{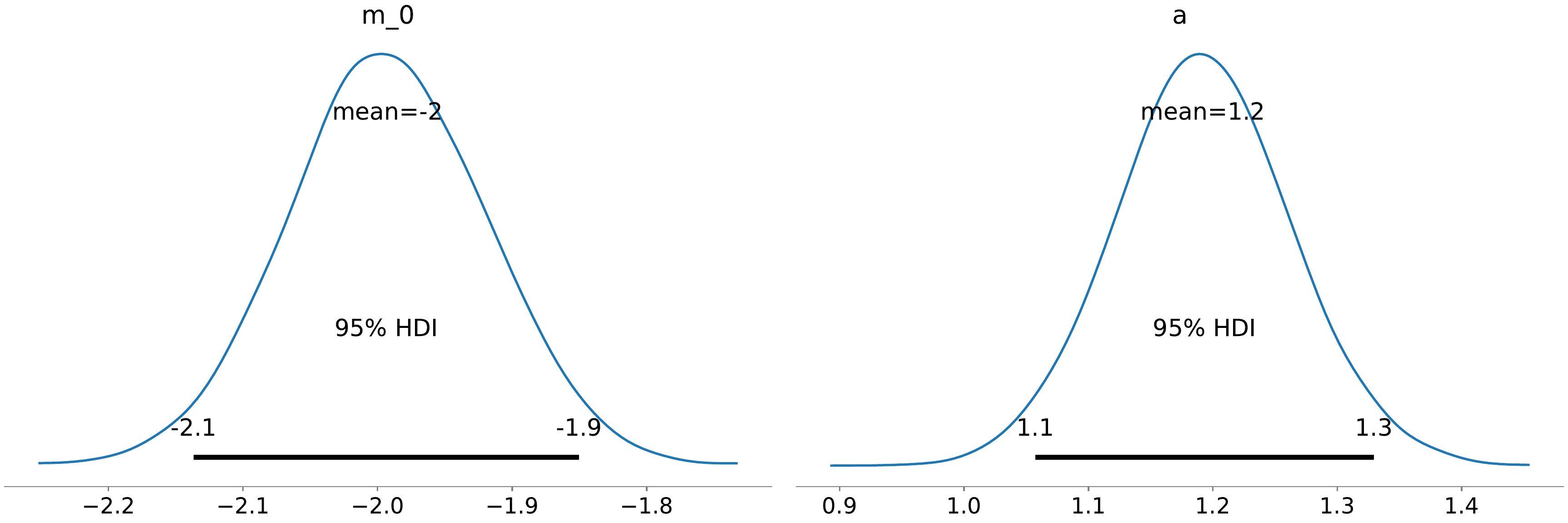} 
	\includegraphics[width=0.48\textwidth,angle=0]{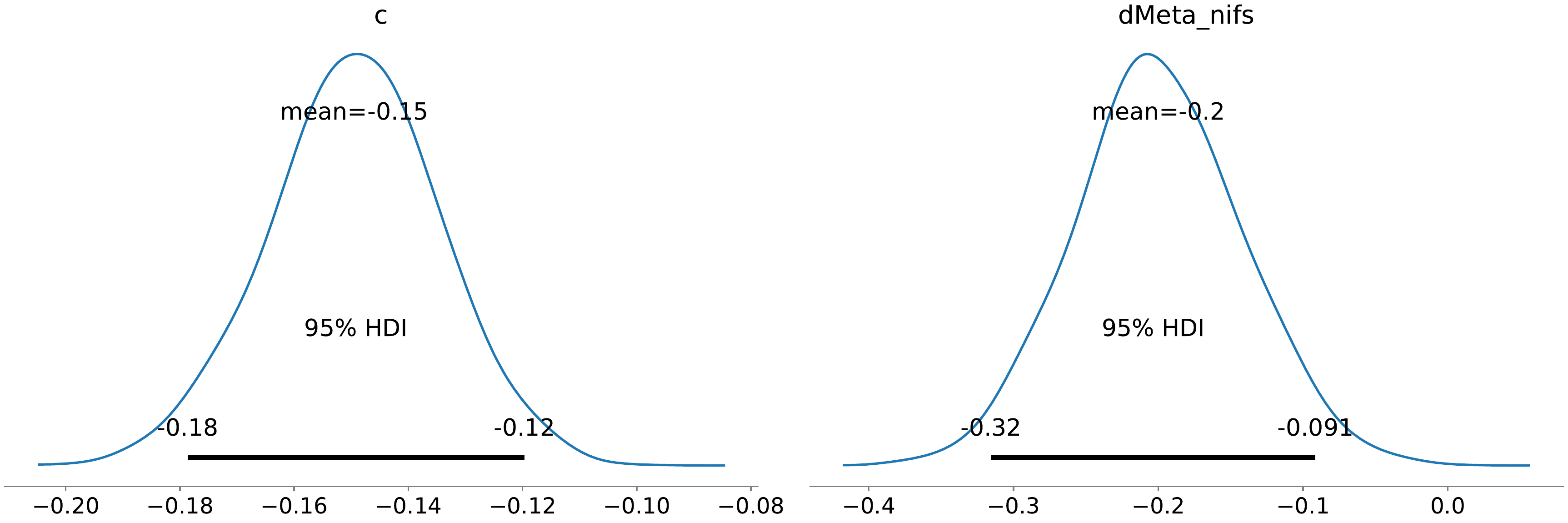} 
    \caption{Similar to Figure \ref{Fig:MetaEWNaI_fits_linear}, but Bayesian fit of the quadratic model. The summary of the posterior distributions and BIC are tabulated in Table \ref{tab:MetaEWNaI_fits}.}
    \label{Fig:MetaEWNaI_fits_quadratic}
\end{figure}

\begin{figure}
	\includegraphics[width=0.45\textwidth,angle=0]{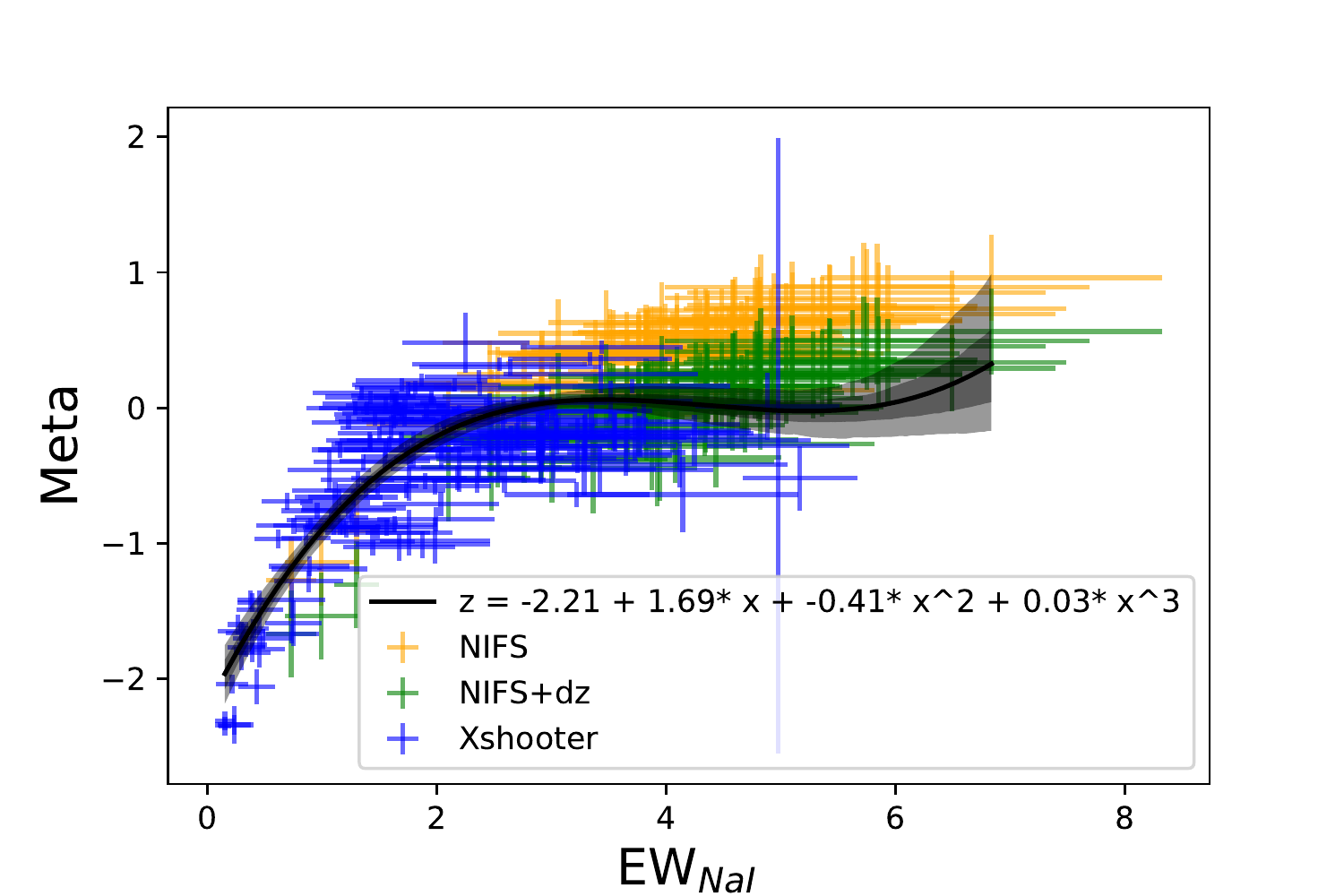} 
	\includegraphics[width=0.47\textwidth,angle=0]{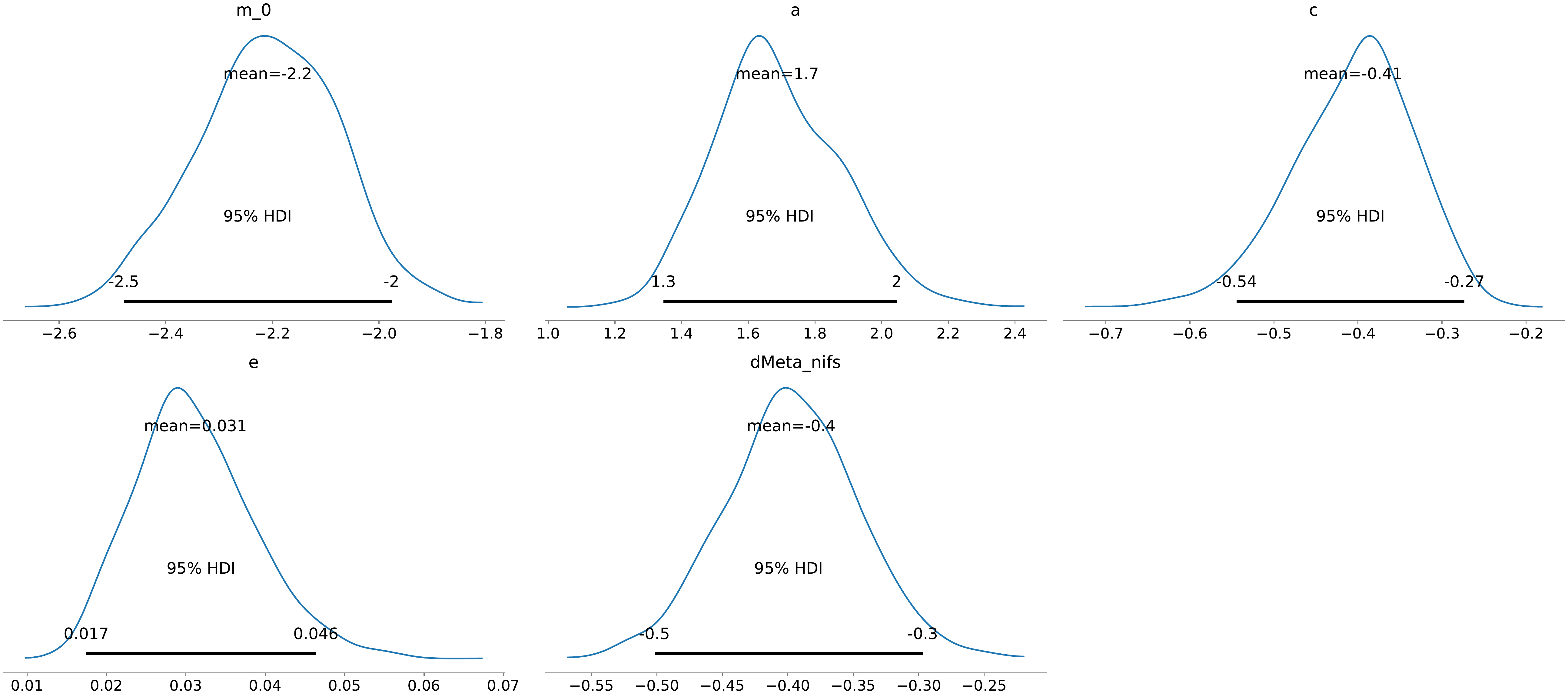} 
    \caption{Similar to Figure \ref{Fig:MetaEWNaI_fits_linear}, but, Bayesian fit of the cubic model. The summary of the posterior distributions and BIC are tabulated in Table \ref{tab:MetaEWNaI_fits}.}
    \label{Fig:MetaEWNaI_fits_cubic}
\end{figure}

Including the posterior distribution of systematics between NIFS and X-shooter data, we can now self consistently address the question of the order of empirical relationship connecting stellar parameter [$Fe/H$] with EWs of NaI. For this analysis, we use the same model framework described in Figure \ref{Fig:KruschkeDiagramLinModel}, but with the Regression formula updated with quadratic and cubic terms of $x$ corresponding to each model. We use all the X-shooter and NIFS data for this analysis. We use the values in Table \ref{tab:dz_posterior} as highly informative priors for $dMeta_{NIFS}$. Figure \ref{Fig:MetaEWNaI_fits_linear}, Figure \ref{Fig:MetaEWNaI_fits_quadratic} and Figure \ref{Fig:MetaEWNaI_fits_cubic} show the linear, quadratic and cubic Bayesian fit of the [$Fe/H$] versus EW$_{NaI}$ relationship, respectively. Table \ref{tab:MetaEWNaI_fits} summarises the posteriors of the coefficients from the fit, as well as the Bayesian Information Criteria (BIC) for each model. BIC penalises for the complexity of the model (degrees of freedom). The cubic model has significantly lower BIC than quadratic ($\Delta$BIS = 69) or linear models ($\Delta$BIC = 170), confirming the frequentist method based results in Section \ref{Metallicity_correlation}.

\begin{table}
\centering
\caption{Model comparison of the linear, quadratic and cubic model in [Fe/H] versus EW$_{NaI}$.}
\label{tab:MetaEWNaI_fits}
\resizebox{0.45\textwidth}{!}{
\begin{tabular}{lllllll} 
   \hline
Model & m$_0$ & $a$ & $c$ & $e$ &$dz_{NIFS}$ & BIC\\	 
\hline		
Linear & --1.013 $\pm$ 0.068 & 0.300 $\pm$ 0.027  & ... & ... & --0.175 $\pm$	0.068 & 500 \\
Quadratic &  --1.993 $\pm$ 0.073 & 1.192 $\pm$ 0.069  & --0.149 $\pm$ 0.015 & ... & --0.200 $\pm$ 0.058  & 399 \\
Cubic & --2.209 $\pm$ 0.130  & 1.687 $\pm$ 0.186 & --0.406 $\pm$ 0.072 & 0.031 $\pm$ 0.008 & --0.397 $\pm$ 0.052 & 330\\
\hline
\end{tabular}
} 
\end{table}

\subsection{Linear model for Teff versus EW$_{CO229}$} \label{LinearModelTeff_CO229}
\begin{figure}
	\includegraphics[width=0.45\textwidth,angle=0]{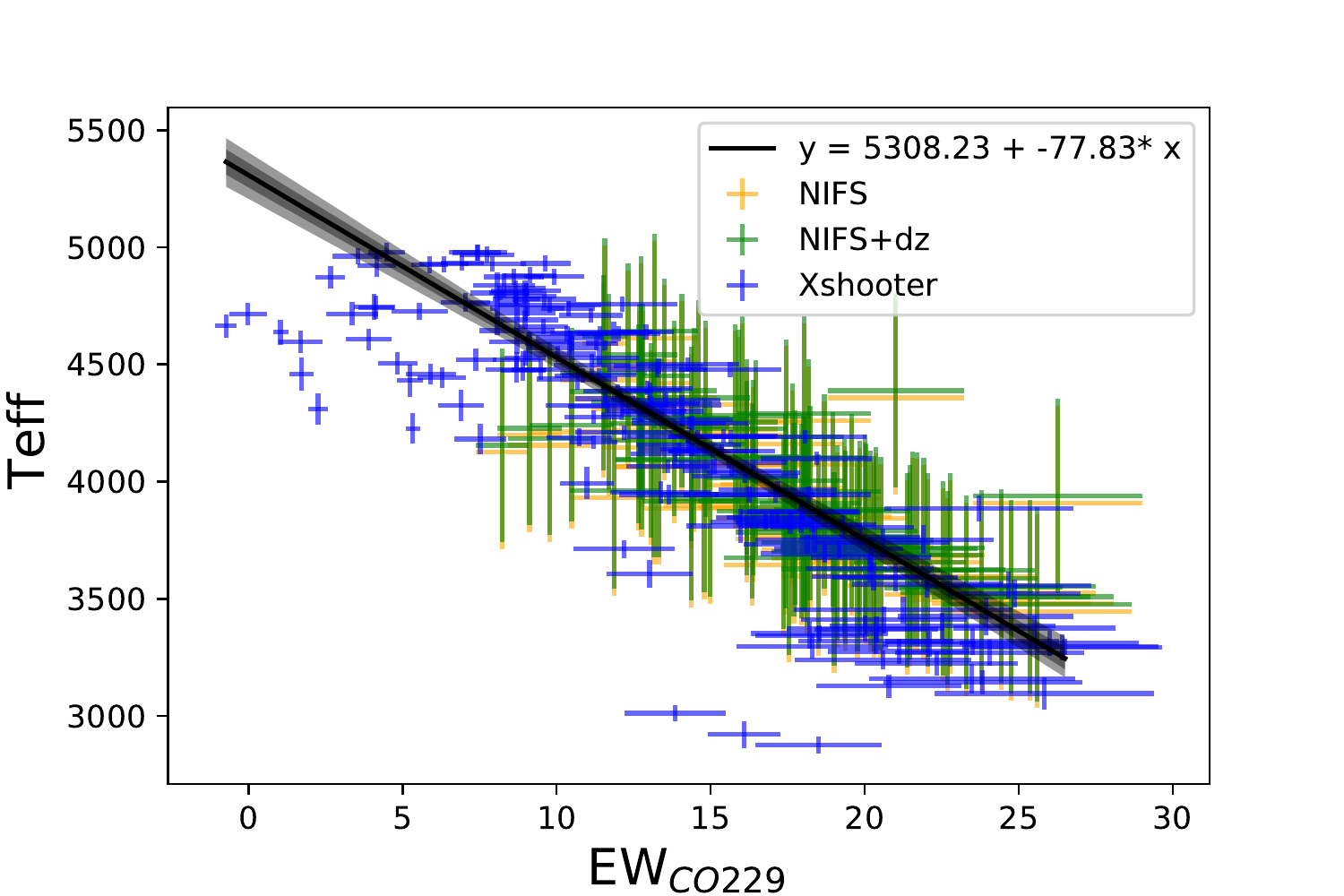} 
	\includegraphics[width=0.46\textwidth,angle=0]{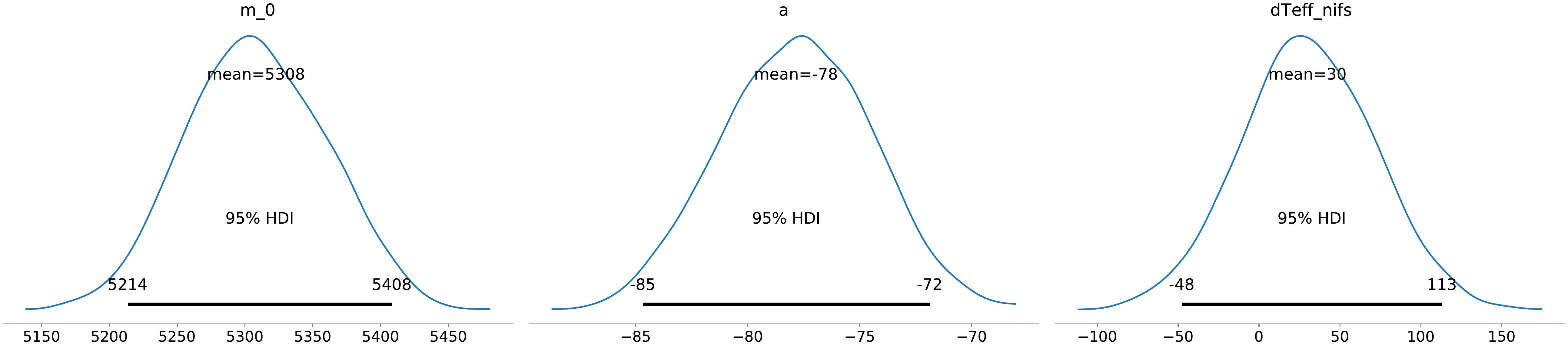} 
    \caption{Bayesian fit of the linear model in $T{_eff}$ versus EW$_{CO229}$ is shown at the top panel. Black curve is the best fit, and the grey shaded regions are  the 1 and 2 sigma intervals of the fitted model. The blue points are the X-shooter data, yellow points are the NIFS data, and green points are the NIFS data offset by the best estimate of the stellar parameter systematic $dTeff_{NIFS}$. The bottom panel displays the posterior distributions of the coefficients in the model. The 95\% Highest Density Interval (HDI) is also marked inside the posterior distributions. The summary of the posterior distributions and BIC are tabulated in Table  \ref{tab:TeffEWCO229_fits}. }
    \label{Fig:TeffEWCO229_fits}
\end{figure}

Just like the linear model in the previous section, the same linear model described in Figure \ref{Fig:KruschkeDiagramLinModel} can be used to fit the linear relationship between Teff and EW$_{CO229}$ as well. We use the values in Table \ref{tab:dz_posterior} as highly informative priors for $dTeff_{NIFS}$. Figure \ref{Fig:TeffEWCO229_fits} shows the linear Bayesian fit of the Teff versus EW$_{CO229}$ relationship. The bottom panel displays the posterior distribution of the group average coefficients. Table \ref{tab:TeffEWCO229_fits} summarises the posteriors of the coefficients from the fit.

\begin{table}
\centering
\caption{Posteriors of the linear model in $T_{eff}$ versus EW$_{CO229}$ relationship.}
\label{tab:TeffEWCO229_fits}
\resizebox{0.45\textwidth}{!}{
\begin{tabular}{lllll} 
   \hline
Model & m$_0$ & $a$ &$dz_{NIFS}$ & BIC\\	 
\hline		
Linear & 5308 $\pm$ 51  & --78 $\pm$ 3.3  & 30 $\pm$ 41 & 3812 \\

\hline
\end{tabular}
} 
\end{table}
\subsection{Metallicity effect on $T_{eff}$--CO229 relation} \label{MetaEffectOnTeffCO229}

\begin{figure}
	\includegraphics[width=0.47\textwidth,angle=0]{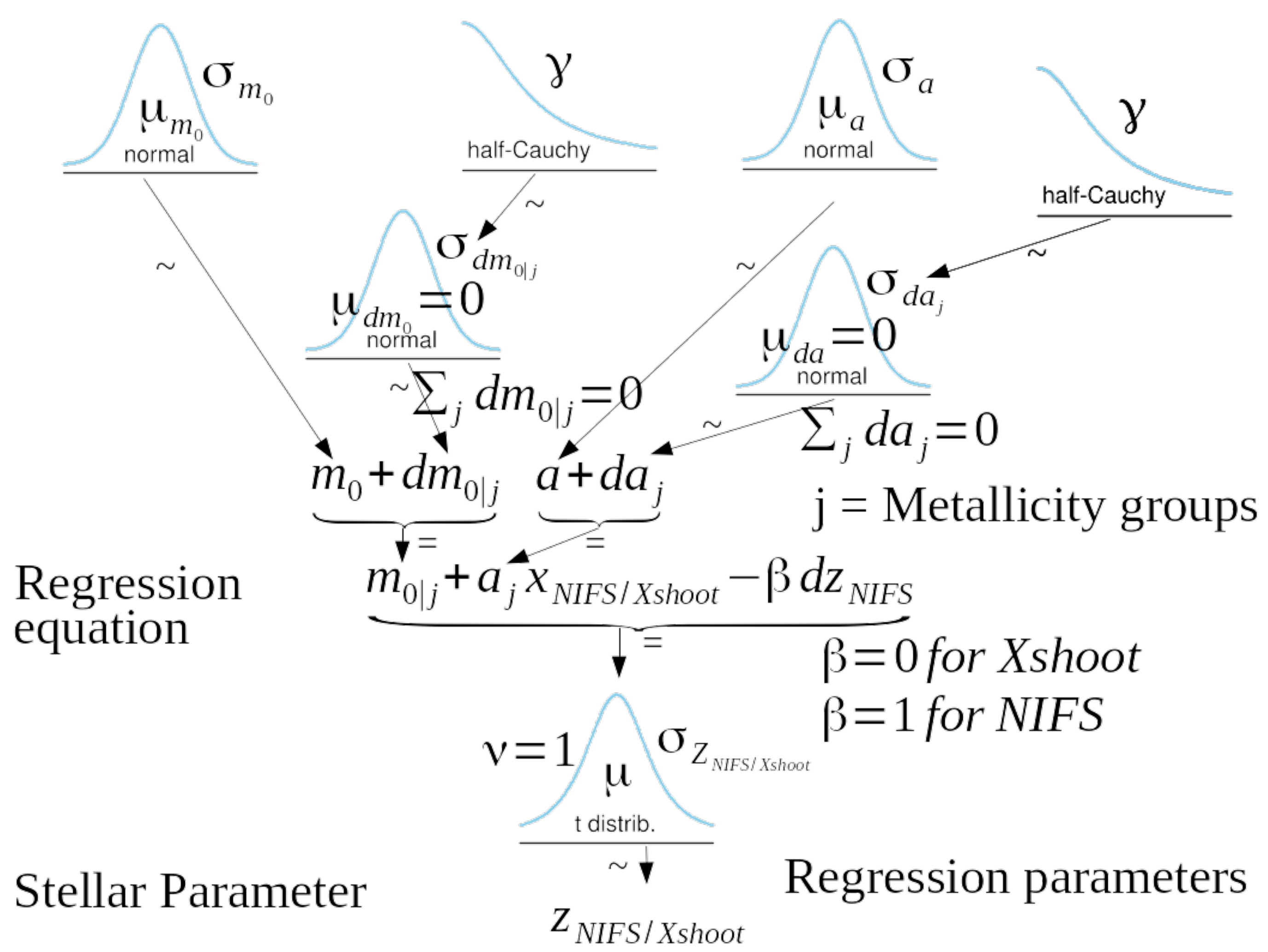}
	\caption{Kruschke style diagram of the Hierarchical Bayesian model to study differences between the Teff versus EW$_{CO229}$ relationship in different metallicity groups. As explained in the text, the coefficients of the linear relationship is hierarchically split into a group average plus delta differences for each metallicity groups. The hyper parameters determine the scatter in these coefficients across the metallicity groups}
	\label{Fig:KruschkeDiagramHBM}
\end{figure}

In order to explore the differences in the linear $T_{eff}$ versus EW$_{CO229}$ relationship for different metallicity group of stars, we developed a Hierarchical Bayesian model. Instead of modelling a heterogeneous set of disjoint metallicity groups, and then comparing the coefficients; Hierarchical Bayesian modelling allows us to simultaneously model different metallicity groups. This enables the model to pool information across the groups while fitting. In some ways, this is the Bayesian equivalent of frequentist MANOVA. Figure \ref{Fig:KruschkeDiagramHBM} shows the Kruschke style diagram of our Hierarchical model. The coefficients of the linear regression equation are modelled as a sum of a group average plus a delta specific to each metallicity group. The sum of all the delta correction to each group is constrained to be equal to zero. The delta correction itself is sampled from a Gaussian distribution with mean zero, and finite sigma. This sigma which represents the scatter in the metallicity group differences is hierarchically sampled from a half-Cauchy distribution with hyper-parameters.

\begin{figure}
	\includegraphics[width=0.47\textwidth,angle=0]{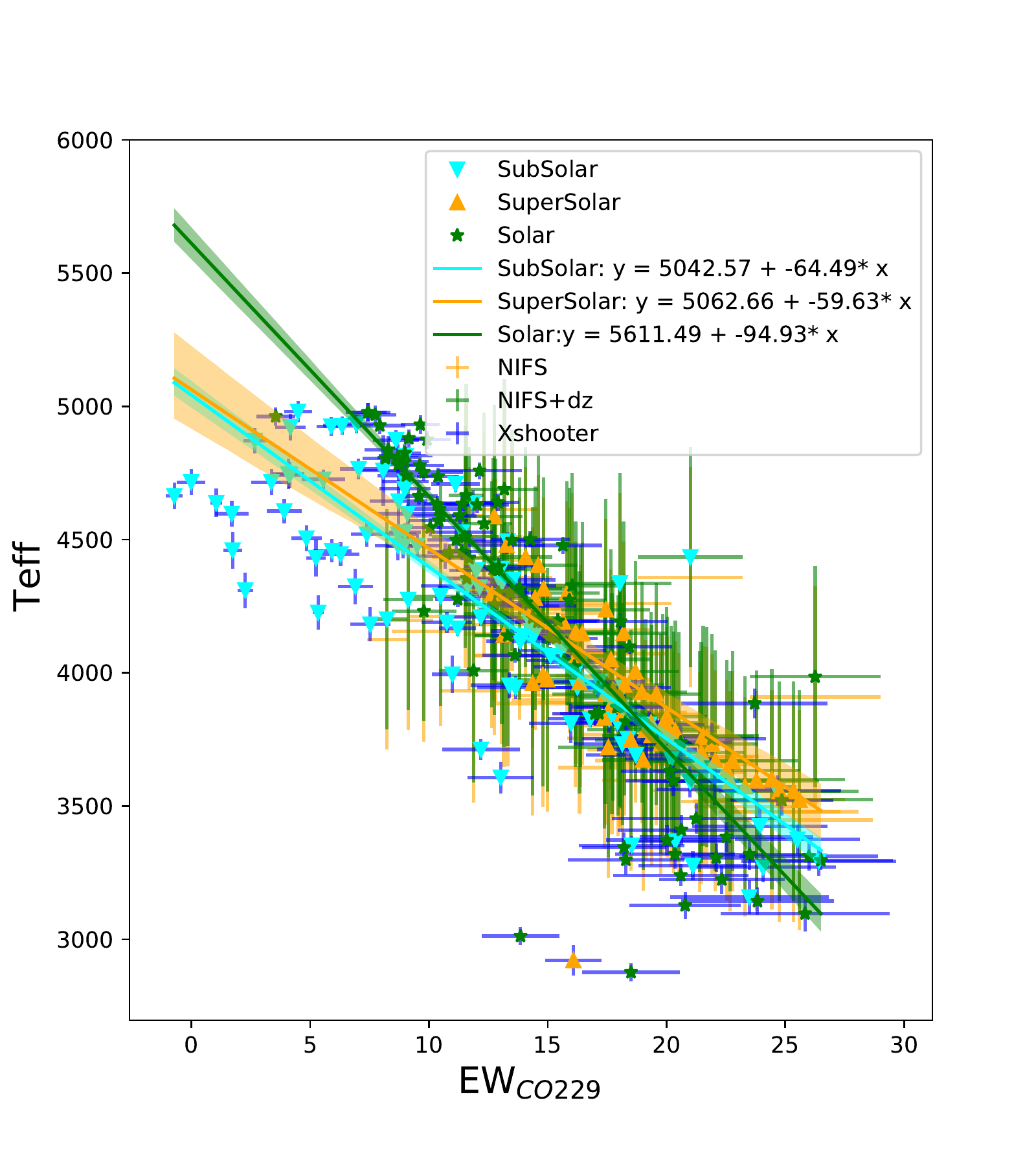} 
    \caption{Metallicity effect in Teff-CO relation shown by grouping data into three separate metallicity groups. The blue error bars are of the X-shooter data, yellow error bars are of the NIFS data, and green drror bars are the NIFS data offset by the best estimate of the stellar parameter systematic $dTeff_{NIFS}$. The cyan colored points show sub-solar metallicity ([Fe/H] $<$ --0.3) group, Orange points label the super-solar metallicity ([Fe/H] $>$ 0.3), and green points label the solar metallicity (+0.3 dex  $>$ [Fe/H] $>$ --0.3 dex) group. The 1 sigma interval of the fitted models for each metallicity group is also shown by the correspondingly colored regions around the best fitted model curves.}
    \label{Fig:Fig:HBM_fitplots}
\end{figure}

\begin{figure}
	\includegraphics[width=0.50\textwidth,angle=0]{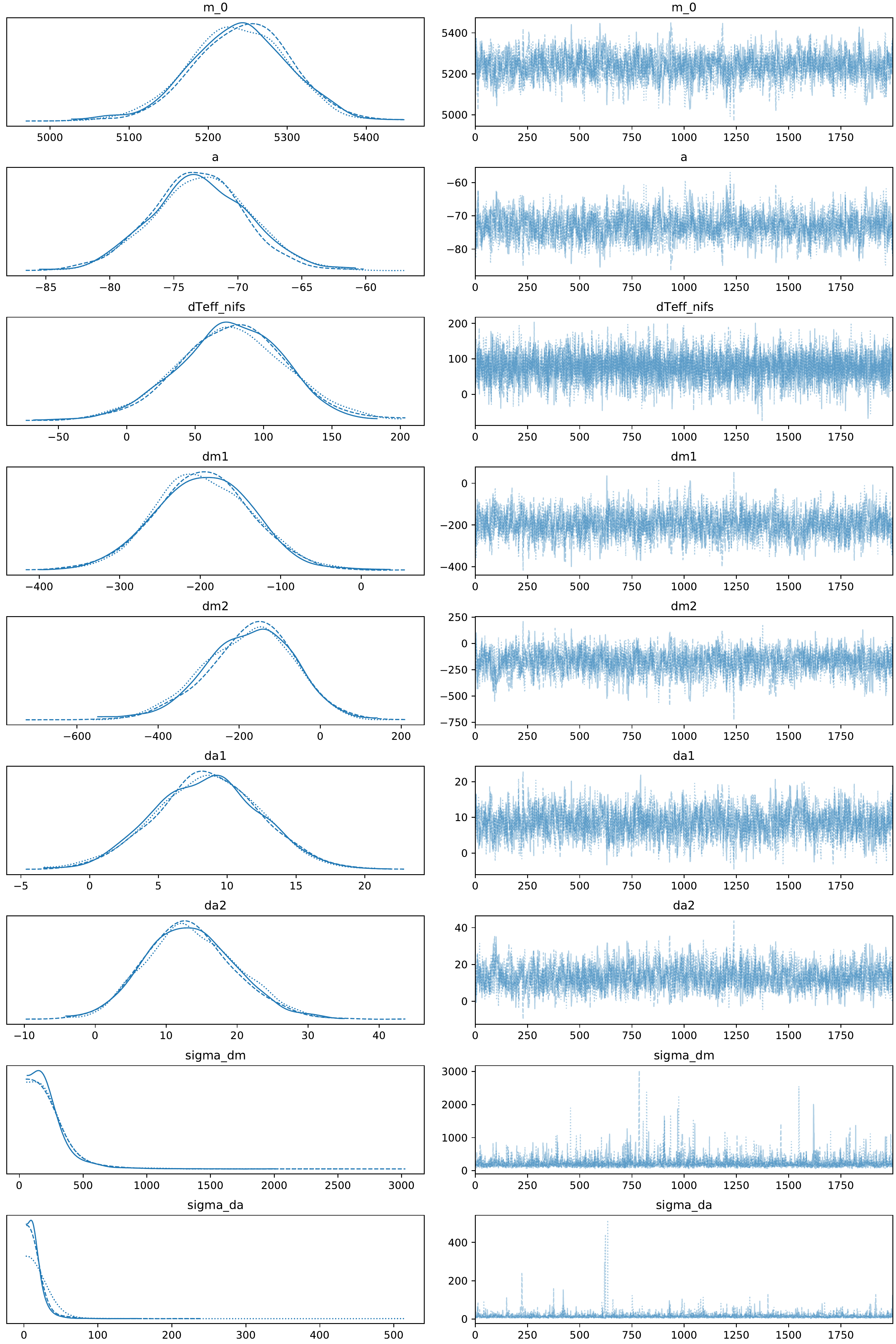} 
	\includegraphics[width=0.5\textwidth,angle=0]{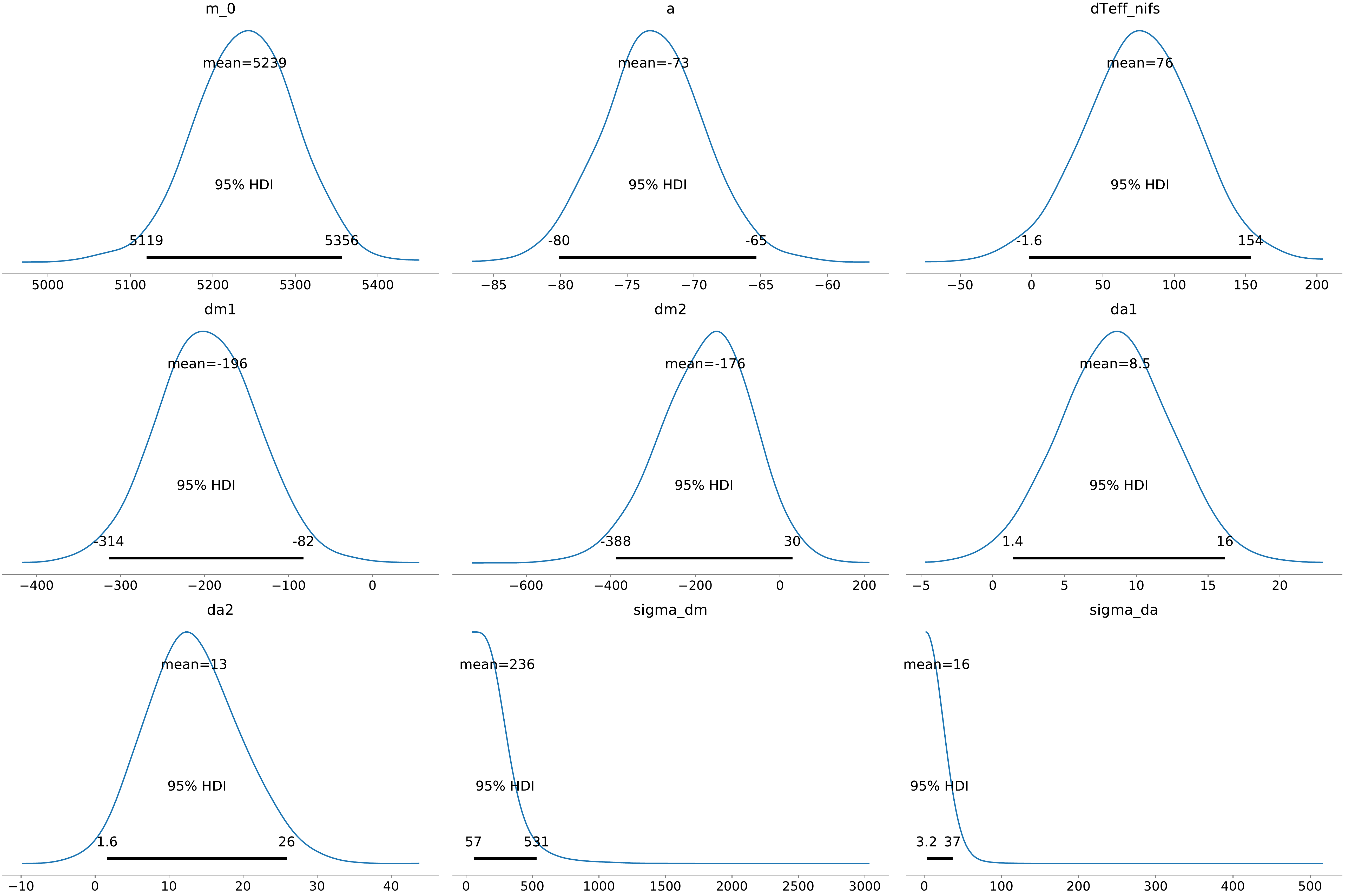} 
    \caption{Top panels shows the trace as well as the posterior distribution of the group average coefficients, as well as the group specific difference to the coefficients of the first two metallicity groups. The third metallicity group's value is given by the constrain that the sum of the group differences should be equal to zero.}
    \label{Fig:HBM_posteriortrace}
\end{figure}

For this analysis we define our metallicity groups to be super-solar ([Fe/H] $>$ +0.3 dex) , solar (+0.3 dex  $>$ [Fe/H] $>$ --0.3 dex), and sub-solar ([Fe/H] $<$ --0.3). Figure \ref{Fig:Fig:HBM_fitplots} shows the three different relations along with their confidence for the three metallicity groups. Figure \ref{Fig:HBM_posteriortrace} shows the traces and the posterior distribution of the variables in the model.
The posterior distributions of the group differences ($dm$ and $da$ terms) show the solar metallicity group's $T_{eff}$ versus EW$_{CO229}$ relationship is significantly different from the sub-solar metallicity group. The difference to the super-solar metallicity group is not as statistically significant. 

\bsp	
\label{lastpage}

\end{document}